\documentclass[12pt]{article}
 \usepackage{hyperref}
 \usepackage{graphicx}
  \usepackage{epsfig}
  \usepackage{amssymb}
  \usepackage{subfigure}
  \usepackage{amsmath} 

\evensidemargin - .5truecm
\allowdisplaybreaks[1]

\begin{document}

\newcommand{\lsim}{\lesssim}
\newcommand{\gsim}{\gtrsim}

\newcommand{\CC}{{\mathbb C}}
\newcommand{\RR}{{\mathbb R}}
\newcommand{\ZZ}{{\mathbb Z}}
\newcommand{\QQ}{{\mathbb Q}}
\newcommand{\NN}{{\mathbb N}}
\newcommand{\beq}{\begin{equation}}
\newcommand{\eeq}{\end{equation}}
\newcommand{\beal}{\begin{align}}
\newcommand{\eeal}{\end{align}}
\newcommand{\nn}{\nonumber}
\newcommand{\bea}{\begin{eqnarray}}
\newcommand{\eea}{\end{eqnarray}}
\newcommand{\ba}{\begin{array}}
\newcommand{\ea}{\end{array}}
\newcommand{\bfig}{\begin{figure}}
\newcommand{\efig}{\end{figure}}
\newcommand{\bc}{\begin{center}}
\newcommand{\ec}{\end{center}}

\newenvironment{appendletterA}
{
  \typeout{ Starting Appendix \thesection }
  \setcounter{section}{0}
  \setcounter{equation}{0}
  \renewcommand{\theequation}{A\arabic{equation}}
 }{
  \typeout{Appendix done}
 }
\newenvironment{appendletterB}
 {
  \typeout{ Starting Appendix \thesection }
  \setcounter{equation}{0}
  \renewcommand{\theequation}{B\arabic{equation}}
 }{
  \typeout{Appendix done}
 }

\begin{titlepage}
\nopagebreak

\renewcommand{\thefootnote}{\fnsymbol{footnote}}
\vskip 1cm

\vspace*{.5cm}
\begin{center}
{\Large \bf 
  Energy-Energy Correlation  in the back-to-back
\\[0.1cm]
 region at N$\bf ^3$LL$\, + \,$NNLO in QCD \\[0.1cm]
}
\end{center}


\par \vspace{1.5mm}
\begin{center}
  {\bf  Ugo Giuseppe\ Aglietti${}^{(a)}$}  and~ {\bf Giancarlo Ferrera${}^{(b)}$}\\

\vspace{5mm}

${}^{(a)}$
Dipartimento di Fisica, Universit\`a di Roma ``La Sapienza'' and\\ INFN, Sezione di Roma
I-00185 Rome, Italy\\\vspace{1mm}

${}^{(b)}$ 
Dipartimento di Fisica, Universit\`a di Milano and\\ INFN, Sezione di Milano,
I-20133 Milan, Italy\\\vspace{1mm}

\end{center}

\vspace{.5cm}


\par 
\begin{center} {\large \bf Abstract} \end{center}
\begin{quote}
\pretolerance 10000

We consider the Energy-Energy Correlation (EEC) function in high-energy electron-positron annihilation to hadrons. 
In the back-to-back (two-jet) region, we perform the all-order resummation of the logarithmically-enhanced 
contributions in QCD perturbation theory up to next-to-next-to-next-to-leading logarithmic (N$^3$LL) accuracy.
Away from the back-to-back region, we consistently combine resummed predictions with the known fixed-order results 
up to next-to-next-to-leading order (NNLO) and we are able to obtain
an accurate fit of the $\mathcal{O}(\alpha_S^3)$ remainder function from the  numerical QCD computation of the full spectrum.
All perturbative terms up to order $\alpha_S^3$ are included
in our calculation and a non-trivial cross-check 
in the back-to-back region
is obtained by comparing the SCET analytic calculation against the corresponding numerical QCD computation.
In particular, the values of the $\mathcal{O}(\alpha_S^3)$ resummation coefficients have been numerically verified.
We regularize the Landau singularity of the QCD coupling within the so-called Minimal Prescription
and we discuss the reduction of the perturbative scale dependence of 
distributions at higher orders, as a means to estimate the corresponding residual perturbative uncertainty.
Finally, after introducing within a dispersive approach non-perturbative power corrections,
we are able to obtain an accurate description of experimental data at LEP and SLC accelerators.
\vskip .2cm

\vfill
\end{quote}
\vspace*{0.5cm}
\begin{center}
{\itshape \large This paper is dedicated to the memory of Stefano Catani,\\\vspace{1mm}
wonderful person, outstanding scientist.}
\end{center}
\vspace*{1cm}
\vspace*{\fill} 

\begin{flushleft}
June 2024
\vspace*{-1cm}
\end{flushleft}
\end{titlepage}

\renewcommand{\thefootnote}{\fnsymbol{footnote}}

\newpage


\section{Introduction}

One of the classical methods to test QCD predictions
and obtain a precise determination of the strong coupling $\alpha_S$
at a reference scale, concerns the analysis of
(infrared-safe) shape variable distributions in high-energy
electron-positron annihilation to hadrons.
In the two-jet region, in which the high-energy hadrons
in final states are collimated into two opposite directions,
the ordinary QCD perturbative expansion does not provide
a good approximation, because its coefficients 
contain large double logarithms of infrared (soft and collinear) origin,
the so-called Sudakov logarithms.
The resummation to all orders in $\alpha_S$ of such
logarithms for shape variables has been formulated in 
the classic paper \cite{Catani:1992ua}.

Among the various shape variables, the Energy-Energy Correlation (EEC)
function\,\cite{Basham:1978bw}  
has received considerable interest over the years
both from the experimental and theoretical 
side.
This function describes the distribution of the 
angular separation (usually called "$\chi$")
of the hard particle pairs in the events
(the precise definition will be given below in Eq.\,(\ref{EEC_def})).
The resummation of Sudakov logarithms in the back-to-back region 
($\chi\to \pi^-$) for the EEC function 
has been achieved 
in full QCD at next-to-leading logarithmic (NLL)%
\,\cite{Collins:1981uk,Collins:1981va,Collins:1985xx,Kodaira:1981nh,Kodaira:1982az},
and at next-to-next-to-leading logarithmic (NNLL)%
\,\cite{deFlorian:2004mp,Tulipant:2017ybb,Kardos:2018kqj} accuracies.
More recently, 
the EEC function has also been analyzed
within the framework of the Soft-Collinear Effective Theory
(SCET) up to next-to-next-to-next-to-leading
logarithmic (N$^3$LL) accuracy\,\cite{Moult:2018jzp,Ebert:2020sfi} and
beyond\,\cite{Duhr:2022yyp}.
The perturbative expansion
of the EEC function also contains large (single)
logarithmic corrections 
in the forward region ($\chi\to 0^+$),
where two or more energetic hadrons
are produced at small angular separations.
However, these effects are physically quite different,
being of hard-collinear nature\,\cite{Dixon:2019uzg};
in this paper we will focus on the back-to-back region.

Away from the endpoints of the angular domain,  the perturbative
series is well behaved, so that 
calculations based on the truncation at a fixed order in $\alpha_S$
are theoretically justified. Since hadron production away from the back-to-back region
has to be accompanied by the radiation of at least one hard recoiling parton,
the leading-order (LO) term for this observable is $\mathcal{O}(\alpha_S)$. 
The LO distribution of the EEC function
has been originally calculated in the late seventies
in Ref.\,\cite{Basham:1978bw}.
The next-to-leading-order (NLO) QCD corrections have been known numerically long ago\,\cite{Richards:1982te,Richards:1983sr} and have been recently computed 
analytically in Ref.\,\cite{Dixon:2018qgp}.   
The next-to-next-to-leading-order (NNLO) correction
has been obtained by numerical Monte Carlo integration of the fully differential cross section for three-jet production in electron-positron annihilation
at NNLO order in QCD
\,\cite{Tulipant:2017ybb,DelDuca:2016csb,DelDuca:2016ily}.
The behavior of the EEC function in the back-to-back region
has been determined analytically at
next-to-next-to-next-to-leading order
(N$^3$LO) in \cite{Ebert:2020sfi}. 

In general, resummed and fixed-order calculations have to be consistently combined
with each other at intermediate values of the angular separation, 
where they are both valid,
in order to obtain accurate QCD predictions on a wide kinematical region.

In this work, we perform the resummation of infrared Sudakov logarithms
in the back-to-back region of the EEC function
up to N$^3$LL accuracy
in QCD,
matching with the corresponding fixed-order results of 
Refs.\,\cite{Basham:1978bw,Dixon:2018qgp,Tulipant:2017ybb} up to  NNLO.
All perturbative terms up to 
N$^3$LO, i.e.\ up to $\mathcal{O}(\alpha_S^3)$, are consistently included\,\footnote{
In the literature this is sometimes referred as N$^3$LL' accuracy.}; 
in particular,	we determine 
the N$^3$LO coefficients of the hard-virtual factor $H(\alpha_S)$ and of the single
logarithmic function $B(\alpha_S)$ of the Sudakov form factor
from the calculation in Ref.\,\cite{Ebert:2020sfi}.
Thanks to the unitarity constraint of the resummation formalism\,\cite{Bozzi:2005wk},
our calculation exactly reproduces, 
after integration over the angular separation variable $\chi$, the corresponding fixed-order result for the total cross section of electron-positron annihilation into hadrons
up to N$^3$LO\,\cite{Gorishnii:1990vf,Surguladze:1990tg}.


The paper is organized as follows.
In Sec.\,\ref{EEC_generic} we define
the EEC observable and consider its standard,
i.e.\ fixed-order perturbative expansion.
In Sec.\,\ref{EEC_back-to-back_resum}
we discuss the QCD impact-parameter ($b$-)space resummation formalism
for the EEC distribution, which has a structure similar to (and simpler than)
the one for the transverse-momentum ($q_T$-)distribution  
in hadron-hadron collisions\,
\cite{Parisi:1979se,Collins:1984kg,Catani:2000vq,Bozzi:2005wk}. 
We also check the consistency,
in the back-to-back region,
between the $\mathcal{O}(\alpha_S^3)$
analytic SCET computation 
against the corresponding numerical computation in full QCD.
In Sec.\,\ref{sec_resum_coef}
we provide explicit formulae for the coefficients
which are needed for the resummation up to 
N$^3$LL+NNLO accuracy (including the hard-virtual coefficients up to N$^3$LO).
In Sec.\,\ref{sec_remainder} we present explicit formulae
for the remainder component of the EEC function,
which is relevant outside the back-to-back region, again at NNLO.
By guessing the functional form for the third-order
remainder function, as well as by using all the available
theoretical information, we are able to obtain a good
fit for this function.
In Sec.\,\ref{sec_scales}
we exhibit and discuss the dependence of our predictions
on auxiliary perturbative scales   at NLL+LO, NNLL+NLO and N$^3$LL+NNLO. 
Such scale dependence is used to
estimate the corresponding perturbative uncertainty of our prediction.
We then extend the perturbative results including  Non-Perturbative (NP) power-behaved QCD contributions\,\cite{Kodaira:1981np,Fiore:1992sa,Nason:1995np,Schindler:2023cww}
through the analytic ``dispersive approach'' of Ref.\,\cite{Dokshitzer:1995qm,Dokshitzer:1999sh}
and we present an illustrative comparison with experimental data from
LEP and SLC accelerators\,\cite{DELPHI:1992qrr,L3:1992btq,OPAL:1991uui,OPAL:1993pnw,SLD:1994idb}.

While we anticipate that 
the N$^3$LL+NNLO prediction provides the best perturbative description of experimental data,
we also remark that there is still a sizable difference
between data and theory, especially in the peak region, where most data are taken.
Conversely, we show that perturbative QCD predictions together with the two-parameters  dispersive model for NP QCD effects provide a good description of experimental data at
the center-of-mass energy of the $Z^0$ resonance. 
Finally, Sec.\,\ref{sec_concl} contains the conclusions of our
analysis.


\section{Energy-Energy Correlation Function}
\label{EEC_generic}

The differential distribution for the Energy-Energy Correlation (EEC)
function in electron-positron annihilation to hadrons is defined as
\beq
\frac{ d \Sigma }{d\cos\chi}
\, = \, 
\sum_{i,j=1}^n 
\int \frac{E_i}{Q} \, \frac{E_j}{Q} \, 
\delta\left( \cos\chi \, - \, \cos\theta_{ij} \right)  
\, d\sigma\,,
\label{EEC_def}
\eeq
where $Q \equiv \sqrt{s}$ is the center-of-mass energy of the colliding leptons,
$n\ge 2$ is the number of hadrons in the (physical) event
(partons in the perturbative QCD calculation) and
$\theta_{ij}$ is the angle between the spatial momenta
$\vec{p}_i$ and $\vec{p}_j$ of hadrons $i$ and $j$, 
respectively ($0\le \theta_{ij} \le \pi$).
Note that self-correlation terms, namely the terms for $i=j$ 
(for which $\theta_{ij} \equiv 0$), are included in the (double) sum.
This distribution characterizes the angular separation
of pairs of hard hadrons in the events.
By integrating over all the relative angles,
the kinematical constraint given by the $\delta$-function
disappears and the total cross section ($\sigma_{\rm tot}$) 
is recovered:
\beq
\int\limits_{-1}^{+1} \frac{ d \Sigma }{d\cos\chi} \, d\cos\chi
\, = \, \int \left( \sum_{i=1}^n \frac{E_i}{Q} \right)^2 \, d \sigma
\, = \, \sigma_{\rm tot}\,,
\eeq
where we have used the relation
\beq
\sum_{i=1}^n E_i \, = \, Q \,.
\eeq
Note that, in order to obtain the total cross section
upon integration, the inclusion of the self-correlation terms
was crucial.

In the center-of-mass frame (where $\vec{P}_{tot}\equiv \sum_{i=1}^n \vec{p}_i=0$),
at lowest order in the expansion in the QCD coupling $\alpha_S$, the final
state consists of a back-to-back quark-antiquark pair
($E_1 = E_2 = Q/2$), implying:
\beq
\label{eq_EEC_lowest_ord}
\frac{1}{\sigma_{0}} \, \frac{ d \Sigma }{d\cos\chi}
\, = \, 
\frac{1}{2} \, \delta(1 - \cos\chi) \, + \, \frac{1}{2} \, \delta(1 + \cos\chi) 
\, + \, \mathcal{O}\left(\alpha_S\right) \, ,
\eeq
where $\sigma_0$ is the total cross section 
in Born approximation for the process $e^+e^-\mapsto \mathrm{hadrons}$.
Let us remark that the factor 2 comes from the fact that the pair $(1,2)$
is counted two times in the double sum above over $i$ and $j$.
The lowest-order distribution then consists of two
peaks of the same strength, at the physical endpoints, $\chi=0$ and $\chi=\pi$\,\footnote{Note the difference with most shape-variable distributions, such as  
for example the thrust, involving a single peak to $\mathcal{O}(\alpha_S^0)$
at a kinematical endpoint.}.

We will mostly consider the back-to-back region, 
\beq
\chi \, \lsim \, \pi
\quad \mathrm{or,\,\,equivalently,} \,\, \pi \, - \, \chi \,  \ll \, 1 .
\eeq
To discuss (perturbative) higher-order corrections, it is convenient to introduce
the unitary variable $z$ defined as:
\beq
z \, \equiv \, \frac{1 \, - \, \cos \chi}{2} \, = \, \sin^2\left( \frac{\chi}{2} \right)\,,
\qquad (0 \le z \le 1)\,.
\eeq
In the back-to-back region, it is also useful to consider the quantity
\beq
1 \, - \, z \, = \, \frac{1 \, + \, \cos\chi}{2}
\, = \, \cos^2\left( \frac{\chi}{2} \right)
\, \simeq \, \left( \frac{\pi \, - \, \chi}{2} \right)^2
\qquad \mathrm{for} \,\,\, \pi \, - \, \chi \, \ll \, 1.
\eeq
In terms of the kinematical variable $z$, the lowest-order EEC function simply reads:
\beq
\label{eq_EEC_lowest_ord_inz}
\frac{1}{\sigma_0} \,\frac{ d \Sigma }{dz}
\, = \, \frac{1}{2} \, 
\delta(1 \, - \, z) \, +  \, \frac{1}{2} \, \delta(z) \, + \, \mathcal{O}\left(\alpha_S\right)\,.
\eeq
The general perturbative QCD expansion of the EEC function, 
normalized to the radiatively-corrected total cross section,
is written as:
\bea
\label{EECfot}
\frac{1}{\sigma_{\rm tot}} \,\frac{ d \Sigma }{dz}
&=& \frac{1}{2} \big[ \delta(1 \, - \, z) \, +  \, \delta(z) \big] 
\, + \, \mathcal{A}(z) \, \frac{\alpha_S}{\pi} 
\, + \, \mathcal{B}(z)  \left( \frac{\alpha_S}{\pi} \right)^2 
\\
&+& 
\,\mathcal{C}(z)  \left( \frac{\alpha_S}{\pi} \right)^3
\, + \, \mathcal{O}\left(\alpha_S^4\right).
\nonumber
\eea
%
The first-order function $\mathcal{A}(z)$ has been evaluated in the late seventies
in \cite{Basham:1978bw};
by including the endpoints contributions\,\cite{Dixon:2019uzg}, it reads:
\bea
\label{eq_Afunc_complete}
\mathcal{A}(z) &=& \frac{4}{3}
\Bigg\{
-  \frac{1}{2} \left[ \frac{\log(1-z)}{1-z} \right]_+
 -  \frac{3}{4} \left[ \frac{1}{1-z} \right]_+
 +  \frac{3}{8} \left[ \frac{1}{z} \right]_+ 
\nonumber\\
&-&  \left( \frac{z_2}{2}  +  \frac{11}{8} \right)  \delta(1-z)
 -  \frac{23}{96}  \delta(z) 
\\
&+&  \frac{1}{8 z^5}
\left[
4 \left( - z^4 - z^3 + 3 z^2 -15 z +9 \right) \log(1-z)
-9 z^4 -6 z^3 - 42 z^2 + 36 z
\right]
\Bigg\}.
\nonumber
\eea
%
The second-order function $\mathcal{B}(z)$ 
has been calculated analytically in  Ref.\,\cite{Dixon:2018qgp},
after the numerical computation in Ref.\,\cite{Richards:1982te,Richards:1983sr}.
Finally, the third-order function $\mathcal{C}(z)$ has been evaluated
only numerically 
in the whole $z$ range in Refs.\,\cite{DelDuca:2016csb,DelDuca:2016ily};
its large $z$ behavior has been evaluated analytically in
Ref.\,\cite{Ebert:2020sfi}.

As can be seen from Eq.(\ref{eq_Afunc_complete})
and from a higher-order analysis,
in the back-to-back region, large logarithms of infrared origin of the form
\beq
\label{logs}
\alpha_S^n \left[ \frac{\ln^k(1-z)}{1-z} \right]_+,
\qquad n = 1,2,3,\cdots, \quad 0\leq k \leq 2n-1,
\eeq
do occur in the perturbative expansion.
The subscript '+' denotes the customary plus regularization
of distributions:
\beq
\int\limits_0^1 \left[ F(z) \right]_+ \varphi(z) \, dz
\, \equiv \,
\int\limits_0^1 F(z) \big[ \varphi(z) \, - \, \varphi(1) \big] \, dz,
\eeq
with $\varphi(z)$ an arbitrary test (i.e.\ smooth) function.
The occurrence of such large logarithmic terms 
in the back-to-back region ($z\to 1^-$)
spoils the convergence of the ordinary, fixed-order perturbative expansion.
To have a reliable QCD prediction, these logarithms  have to be resummed
to all orders of $\alpha_S$;
for consistency reasons, as we are going to show,
one has also to include all the contribution proportional 
to $\delta(1-z)$.


\section{Resummation of Energy-Energy Correlation in the back-to-back region}
\label{EEC_back-to-back_resum}

The differential distribution for the EEC function in Eq.\,(\ref{EECfot})
is decomposed as:
\beq
\label{EEC-dec}
\frac{1}{\sigma_{\rm tot}} \, \frac{ d \Sigma }{dz}
\, = \,
\frac{1}{\sigma_{\rm tot}} \, \frac{ d \Sigma_{(\rm res.)} }{dz}
\, + \, 
\frac{1}{\sigma_{\rm tot}} \, \frac{ d \Sigma_{(\rm fin.)} }{dz}\,.
\eeq
The first term on the right-hand side of Eq.(\ref{EEC-dec}) 
is the {\itshape resummed} $(\rm res.)$ component, 
containing all the logarithmically-enhanced
contributions at large $z$ (see Eq.(\ref{logs})),
to be resummed to all orders (i.e.\ for any $n=1,2,3,\cdots$), together with the $\delta(1-z)$ contributions, 
while the second term, the {\itshape finite} $(\rm fin.)$ component, free from such contributions, can be computed by standard, fixed-order
perturbation theory.


\subsection{Resummed term}

In order to consistently take into account the kinematics constraint of transverse-momentum conservation in multiple parton emissions,
the resummation program has to be carried out in the impact-parameter space
or $b$-space ($b$ is the conjugated variable to $Q\sqrt{1-z}$)\,\cite{Parisi:1979se},
where the back-to-back region $1-z\ll 1$
corresponds to the region $b Q\gg 1$.
The EEC distribution in the physical $z$-space is then recovered
by performing an inverse Fourier-Bessel transformation with respect to the impact parameter $b$
\cite{Collins:1981uk,Collins:1981va,Collins:1985xx,Kodaira:1981nh,Kodaira:1982az}:
\beq
\label{EEC-b}
\frac{1}{\sigma_{\rm tot}} \, \frac{ d \Sigma_{\rm (res.)} }{dz}
\, = \, \frac{1}{2} \, H\left(\alpha_S\right)
\int\limits_0^\infty  d(Q b) \, \frac{Q b}{2} \,
J_0\left( \sqrt{1 - z} \, Qb \right) 
 \, S\left(Q,b\right) \, ,
\eeq
where $J_0(x)$ is the Bessel function of first kind $J_\nu(x)$
with zero index, $\nu=0$.
According to Eq.(\ref{EEC-b}),
the resummed component of the EEC distribution has been factorized, 
in $b$-space,
in two factors, namely $H(\alpha_S)$ and $S(Q,b)$.
The function $H\left(\alpha_S\right)$ in Eq.(\ref{EEC-b}) is a $b$-independent, hard factor, including hard-virtual contributions at a scale $q \sim Q$;
it contains all the terms that behave as constants in the limit $b\to\infty$,
which correspond, in a minimal factorization scheme,
to corrections proportional to $\delta(1-z)$ in the physical (angle) space.
The function $S\left(Q,b\right)$ in Eq.(\ref{EEC-b}) is the QCD Sudakov form factor, which resums to all orders, in $b$-space, the large logarithmic corrections of the type
$\alpha_S^n\ln^k(Q^2 b^2)$ ($n=1,2,3\dots$, $1\leq k\leq 2n$).
The latter are divergent when $b\to+\infty$ and correspond, in physical space, to the
infrared logarithms in Eq.(\ref{logs}).
Note that the (perturbative) factorization of constants and logarithmic
terms in Eq.(\ref{EEC-b}) involves some degree of arbitrariness, 
since the argument of the large logarithms can
always be rescaled as: 
\beq
\label{LL}
\ln \left( \frac{Q^2 \, b^2}{b_0^2} \right) 
\, = \, 
\ln \left(\frac{\mu_Q^2 \, b^2}{b_0^2} \right) 
\, + \, \ln\left( \frac{Q^2}{\mu_Q^2} \right),
\eeq
where we have introduced the coefficient $b_0 \equiv 2 \exp\left(-\gamma_E\right) \simeq 1.123$
($\gamma_E = 0.5772\cdots$  is the Euler-Mascheroni constant)%
\,\footnote{The $b_0$ coefficient has a kinematical origin; its insertion has the sole purpose of simplifying the algebraic expression of $S(Q,b)$.}. 
The rescaling in Eq.\,(\ref{LL}) is governed by the arbitrary
(but independent of $b$) energy scale
$\mu_Q$,  called {\itshape resummation} scale\,\cite{Bozzi:2005wk}.
This scale has to be considered of the order of the hard scale,
$\mu_Q \sim Q$, so that $\ln(Q^2/\mu_Q^2)$ is a constant term $\mathcal{O}(1)$;
in other words, it is assumed not to be a large logarithm.
That implies, in particular, that the $\mu_Q$-dependent terms
have not to be exponentiated in $S(Q,b)$, but rather factorized in 
$H(\alpha_S)$, which therefore explicitly depends on $\mu_Q$.  
The general large (logarithmic) expansion parameter is thus:
\beq
\label{lpar}
L \, \equiv \, \ln \left( \frac{\mu_Q^2 \, b^2}{b_0^2} \right)\,
\eeq
with $L  \gg 1$ for $b \gg b_0/\mu_Q \sim 1/Q$.
Let us notice that
the role played by the resummation scale $\mu_Q$ in the resummation formalism
is analogous to the role played by the renormalization 
scale $\mu_R$ in the context of renormalization.
If evaluated exactly, i.e.\ by formally including  all the perturbative orders, 
the resummed cross section, Eq.\,(\ref{EEC-b}),
does not depend on $\mu_Q$.
On the contrary, when evaluated approximately,
i.e.\ at some level of logarithmic accuracy,
the resummed cross section exhibits a residual
dependence on $\mu_Q$.
%
We choose the central or reference value of the resummation scale to be equal to the hard scale,
$\mu_Q = Q$. 
Actually, conventional variations of $\mu_Q$ around its central value can be used to
estimate the uncertainty coming from yet uncalculated 
higher-order logarithmic terms.

The resummation of the logarithmic contributions is achieved by showing that the Sudakov form factor can be expressed in the following exponential
form\,\cite{Collins:1981uk,Kodaira:1981nh,Bozzi:2005wk}:
\beq
\label{sud}
S\left(Q,b\right) \, = \,
\exp\Bigg\{ -\int\limits_{b_0^2/b^2}^{\mu_Q^2} \frac{dq^2}{q^2}
\left[
A\left(\alpha_S(q^2)\right) \ln\left(\frac{Q^2}{q^2}\right)
\, + \, B\left( \alpha_S(q^2) \right) 
\right] 
\Bigg\}.
\eeq
The double-logarithmic function $A(\alpha_S)$ describes the effects of both soft and collinear parton
emission off the primary partons in the process (a light quark-antiquark pair in our case), 
while the single-logarithmic function $B(\alpha_S)$ describes the effect of 
soft wide-angle radiation or hard-collinear radiation at scales 
$1/b \lesssim q \lesssim Q$ 
(in the relevant kinematic region for resummation,
$b \gg 1/Q$).
Since the EEC function is an infrared-safe observable,
soft-virtual cancellation suppresses both soft and collinear contributions
at energy/momentum scales $q \lesssim 1/b$.
The functions $A(\alpha_S)$, $B(\alpha_S)$ and $H\left(\alpha_S\right)$ possess ordinary  (i.e.\ without logarithmic coefficients)
perturbative expansions in powers of $\alpha_S$ of the form:
\bea
A(\alpha_S) &= & \sum_{n=1}^\infty A_n \,  \left( \frac{\alpha_S}{\pi} \right)^n\,,\label{eqA}\\
B(\alpha_S) & = & \sum_{n=1}^\infty B_n \,  \left( \frac{\alpha_S}{\pi} \right)^n\,,\label{eqB}\\
H(\alpha_S) & = & 1 \, + \,\sum_{n=1}^\infty H_n \,  \left( \frac{\alpha_S}{\pi} \right)^n\,\label{eqH}.
\eea
The integral on the right hand side (r.h.s.) of Eq.\,(\ref{sud}) can be explicitly evaluated order-by-order in $\alpha_S$ by using the iterative solution
of the renormalization group equation for the QCD coupling:
\beq
\label{beta}
\frac{d \ln \alpha_S(\mu_R^2)}{d\ln\mu_R^2} \, \equiv \, \beta(\alpha_S)
\, = \, - \, \beta_0 \, \frac{\alpha_S}{\pi} \, - \, \beta_1  \left( \frac{\alpha_S}{\pi} \right)^2
\, - \, \beta_2  \left( \frac{\alpha_S}{\pi} \right)^3 
\, - \, \beta_3  \left( \frac{\alpha_S}{\pi} \right)^4 \, - \, \cdots\,,
\eeq
where  $\mu_R$ is the renormalization scale.

The solution of Eq.(\ref{beta}) up to N$^3$LO is given by:
\bea
\label{alphas}
\alpha_S(q^2)&=&\frac{\alpha_S}{l}
\left\{
1 \, - \, \frac{\alpha_S}{\pi l} \frac{\beta_1}{\beta_0} \ln l
+ \left(\frac{\alpha_S}{\pi l}\right)^2 \left[ \frac{\beta_1^2}{\beta_0^2} (\ln^2 l-\ln l+l-1) - \frac{\beta_2}{\beta_0} (l-1) \right]\right.
\nn\\
&+& \left(\frac{\alpha_S}{\pi l}\right)^3   \left[ \frac{\beta_1^3}{\beta_0^3} \left(-\ln^3l +\frac52 \ln^2l-2(l-1)\ln l -\frac{(l-1)^2}2 \right) \right.\nn\\
  && ~~~~~~~~~+ \left.\left.\frac{\beta_1 \beta_2}{\beta_0^2} \big( (l-1) l + (2l-3) \ln l\big) + \frac{\beta_3}{\beta_0} \frac{(1-l^2)}2\right]
+\mathcal{O}(\alpha_S^4)
\right\},
\eea
with $\alpha_S \equiv \alpha_S(\mu_R^2)$ and $l \equiv 1 + \beta_0 \alpha_S/\pi \ln(q^2/\mu_R^2)$.
  
After the analytic integration in the exponent in Eq.\,(\ref{sud}), 
the Sudakov form factor  can be recast in the following form\,\cite{Catani:2000vq,deFlorian:2004mp}:
\beq
\label{sud2}
S(Q,b) \, = \,
\exp\left\{ L \, g_1(\lambda) \, + \, g_2(\lambda) 
\, + \, \frac{\alpha_S}{\pi} g_3(\lambda) 
\, + \, \left(\frac{\alpha_S}{\pi}\right)^2g_4(\lambda)
\, + \, 
\sum_{n=5}^{+\infty}\left(\frac{\alpha_S}{\pi}\right)^{n-2}g_n(\lambda)\right\}\,,
\eeq
where the variable
\beq
\lambda \, \equiv \, \frac{\alpha_S}{\pi} \, \beta_0 \, L \,,
\eeq
is assumed to be $\mathcal{O}(1)$.
The {\itshape exponent} on the r.h.s.\ of Eq.\,(\ref{sud2}) has thus a customary perturbative expansion in powers of $\alpha_S$,
with $\lambda$-dependent coefficients.
The truncation of such (function) series at a given order resums an infinite series of
logarithmic corrections.  
The leading-logarithmic (LL) approximation is provided by the function 
$g_1(\lambda)$, the NLL approximation requires also the inclusion of the function
$g_2(\lambda)$,  the NNLL and N$^3$LL approximations
require also the functions $g_3(\lambda)$ 
and $g_4(\lambda)$ respectively, and so on. The resummation
functions $g_n(\lambda)$, with $1\leq n\leq 4$, have the following
explicit expressions:
\bea
g_1(\lambda) =  \frac{A_1}{\beta_0} \, \frac{ \lambda+\ln(1-\lambda)}{ \lambda}\,,
\label{g1}
\eea
\bea
\label{g2}
g_2(\lambda)  &=& 
-  \frac{A_2}{\beta_0^2}  \left[\frac{\lambda}{1-\lambda}+\ln   (1-\lambda) \right]
 +  \frac{{B}_1 }{\beta_0} \ln (1-\lambda)  
\nonumber\\	
&+&  \frac{A_1 \beta_1}{\beta_0^3} \left[\frac{\lambda}{1-\lambda}+\frac{1}{2} \ln^2(1-\lambda)+\frac{\ln(1-\lambda)}{1-\lambda}\right]  
\nonumber\\
&+&  \frac{A_1}{\beta_0} \left[   \frac{\lambda}{1-\lambda}+ \ln(1-\lambda) \right] 
\ln\left( \frac{\mu_Q^2}{\mu_R^2} \right)\,,
\eea
\bea
\label{g3}
g_3(\lambda) &=& 
-  \frac{A_3 }{2  \beta_0^2}  \frac{\lambda^2}{(1-\lambda)^2}
 -  \frac{{B}_2}{\beta_0}  \frac{ \lambda}{ 1-\lambda}
 -  \frac{A_2  \beta_1}{2  \beta_0^3}  
\frac{ \lambda (2-3 \lambda)  +   2 (1-2 \lambda) \ln(1-\lambda)}{(1-\lambda)^2}  
\nonumber\\   
&+&  \frac{A_1  \beta_2 }{2  \beta_0^3} 
\left[
\lambda  \frac{2-3 \lambda}{(1-\lambda)^2}  +  2 \ln(1-\lambda) \right]  
 +  \frac{{B}_1  \beta_1}{\beta_0^2} 
 \frac{ \lambda+\ln (1-\lambda) }{ 1-\lambda}  
\nonumber\\
&+&  \frac{A_1  \beta_1^2}{2  \beta_0^4}  
\frac{ \lambda^2  +  (1-2\lambda) \ln^2(1-\lambda) 
 +  2 \lambda (1-\lambda)  \ln (1-\lambda) }{ (1-\lambda)^2}  
\nonumber\\
&-&  \frac{A_1}{2}  \frac{ \lambda^2}{(1-\lambda)^2}  
\ln^2\left(\frac{\mu_Q^2}{\mu_R^2}\right)
 +  \Bigg[ 
\frac{A_2}{\beta_0}  \frac{ \lambda^2}{ (1-\lambda)^2}
 +  {B}_1  \frac{\lambda}{1-\lambda}  \nn
\\
&+&  \frac{A_1  \beta_1}{\beta_0^2} 
\frac{ (1 - 2 \lambda) \ln (1 - \lambda)  +  \lambda (1 - \lambda) }{
(1-\lambda)^2} 
\Bigg] 
\ln\left(\frac{\mu_Q^2}{\mu_R^2}\right)\,,
\eea
\bea
\label{g4}
g_4(\lambda) &=&
-  \frac{A_4}{6  \beta_0^2}  \lambda^2 \frac{3-\lambda}{ (1-\lambda)^3}
 -  \frac{{B}_3}{2  \beta_0}  \lambda  \frac{2-\lambda}{ (1-\lambda)^2}
 +  \frac{{B}_2  \beta_1}{2  \beta_0^2}  \frac{ 2 \ln
   (1-\lambda)+ \lambda(2-\lambda) }{ (1-\lambda)^2}  
\nonumber\\
&-&  \frac{A_3  \beta_1}{12  \beta_0^3}  
\frac{ \lambda \left(5 \lambda^2-15\lambda+6\right)  +  6(1-3 \lambda) \ln
   (1-\lambda)}{ (1-\lambda)^3} 
 -  \frac{2 A_2  \beta_2}{3  \beta_0^3}  \frac{\lambda^3}{ (1-\lambda)^3}  
\nonumber\\
&+&  \frac{A_2  \beta_1^2}{12  \beta_0^4}  
\frac{ \lambda \left(11 \lambda^2 - 9
   \lambda+6\right)  +  6 (1-3 \lambda) \ln^2(1-\lambda)  +  6 (1-\lambda) \ln
   (1-\lambda)}{(1-\lambda)^3}  
\nn\\
&+&  \frac{A_1  \beta_3 }{12  \beta_0^3}
\left[
\lambda \frac{7 \lambda^2-15\lambda+6}{(1-\lambda)^3}  +  6 \ln(1-\lambda)
\right]  
\nn\\
&-&  \frac{A_1  \beta_1^3}{6  \beta_0^5}  
\frac{ \lambda^3 + 3 \lambda^2 (1+\lambda)  \ln(1-\lambda)
 +  3 \lambda \ln^2(1-\lambda)
 +  (1-3\lambda) \ln^3(1-\lambda) }{(1-\lambda)^3}  
\nonumber\\
&-&  \frac{A_1  \beta_1  \beta_2 }{12  \beta_0^4}  
\frac{\lambda \left(5 \lambda^2  -  15 \lambda+6\right) 
 -  6 \left(2 \lambda^3-2\lambda^2+3 \lambda-1\right) \ln(1-\lambda)
}{(1-\lambda)^3}  
\nn\\
&-&  \frac{{B}_1  \beta_2}{2  \beta_0^2}  \frac{ \lambda^2}{(1-\lambda)^2}
 +  \frac{{B}_1  \beta_1^2}{2  \beta_0^3} 
 \frac{\lambda^2 - \ln ^2(1-\lambda)}{ (1-\lambda)^2} 
 +  \frac{A_1  \beta_0}{6}  \lambda^2  \frac{3-\lambda}{(1-\lambda)^3}  \ln^3\left(\frac{\mu_Q^2}{\mu_R^2}\right)  
\nonumber\\
&-&  \Bigg[ 
\frac{A_2}{2}  \lambda^2 \frac{3-\lambda}{(1-\lambda)^3} 
 +  \frac{A_1  \beta_1}{2  \beta_0}  
\frac{ \lambda + (1-3 \lambda) \ln(1-\lambda) }{ (1-\lambda)^3} 
 +  \frac{{B}_1  \beta_0}{2}  \lambda  \frac{2-\lambda}{(1-\lambda)^2} 
\Bigg]   
\ln^2\left(\frac{\mu_Q^2}{\mu_R^2}\right)  
\nonumber\\
&+&  \Bigg[ 
\frac{A_3}{2  \beta_0}  
\lambda^2 \frac{3-\lambda}{ (1-\lambda)^3} 
 +  {B}_2  \lambda \frac{2-\lambda}{(1-\lambda)^2} 
 +  \frac{A_2  \beta_1}{2  \beta_0^2}  
\frac{2(1-3\lambda) \ln (1-\lambda)+\lambda
\left(\lambda^2-3 \lambda+2\right)}{ (1-\lambda)^3}  
\nn\\&+&  \frac{A_1  \beta_2}{2  \beta_0^2} 
\lambda^2 \frac{1+\lambda}{ (1-\lambda)^3} 
 -  \frac{A_1  \beta_1^2}{2  \beta_0^3}  
\frac{ \lambda^2 (1+\lambda)  + (1-3 \lambda)
   \ln^2(1-\lambda) + 2 \lambda \ln(1-\lambda) }{ (1-\lambda)^3}  
\nonumber\\
&-&  \frac{{B}_1  \beta_1}{\beta_0}  \frac{ \ln(1-\lambda) }{(1-\lambda)^2}
\Bigg] \ln\left(\frac{\mu_Q^2}{\mu_R^2}\right)\,.
\eea
%
The following remarks are in order.
\begin{enumerate}
\item
The expression of the Sudakov form factor given in Eq.\,(\ref{sud}) only has a formal meaning, as it involves integration
over the (non-integrable) Landau singularity of the running coupling when $b > b_0 /\Lambda_{QCD}$, where $\Lambda_{QCD}$ is the QCD scale
(at lowest order the singularity of the coupling is a simple pole as
$\alpha_S(q^2) \sim \pi/\beta_0/\ln(q^2/\Lambda_{QCD}^2)$).
This singularity manifests itself
in singularities of the $g_n(\lambda)$ functions at the point $\lambda=1$
(with increasing strength with the function order $n$)
\footnote{
A similar dynamical mechanism occurs in the 
perturbative analysis of resonances at finite
volume in quantum mechanics \cite{Aglietti:2019ltw,Aglietti:2022fhy}.
},
which corresponds to the value of the impact parameter
$b=b_L= b_0/Q \,\exp\{\pi/(2 \beta_0 \alpha_S)\} \sim 1/\Lambda_{QCD}$.
That implies some prescription  is needed
to regulate the Landau singularity for an {\it effective} evaluation of $S(Q,b)$ 
also in the short-distance (perturbative) region $b\lsim b_L$.
In our numerical study, we have regularized the Landau singularity using the so-called Minimal Prescription\,\cite{Catani:1996yz,Laenen:2000de,Kulesza:2002rh}, that is deforming the integration contour in the complex $b$ space. 
\item
Let us note that the logarithmic variable $L$ defined Eq.\,(\ref{lpar}),
{\it diverges} in the limit $b \to 0^+$. As a consequence also
 the exponent of the form factor in  Eqs.(\ref{sud},\ref{sud2}) 
diverges in such limit. However the limit $b \to 0^+$  corresponds
to the total cross section and resummation effects are not justified. We thus
impose the so-called {\it unitarity constraint}\,\cite{Bozzi:2005wk}
by replacing
everywhere in the form factor the logarithmic variable $L$ with the variable
\beq
\label{Ltilde}
\widetilde{L} \, \equiv \, \log\left(\frac{\mu_Q^2 b^2}{b_0^2}\,+\,1 \right).
\eeq
$\widetilde{L}$ indeed vanishes in the limit $b \to 0^+$ (the form
factor is thus equal to unity)
and, at the same time,
has the same long-distance ($b \to + \infty$) behavior  as $L$ (apart from power corrections $\sim 1/b^2$).
The unitarity constraint avoids the introduction of (unjustified)
resummation contributions in the small-$b$ region,
reducing  perturbative
uncertainties at intermediate values of  $z$, and allows us to
exactly recover the fixed-order total cross section  upon integration over $z$.
\item
The functions $g_n(\lambda)$, for $n\geq 2$, do explicitly depend on the renormalization scale $\mu_R$ and on the resummation scale $\mu_Q$.
The $\mu_R$ dependence is exactly canceled, order-by-order in the perturbation expansion of the exponent of
the form factor in Eq.\,(\ref{sud2}), by the renormalization scale dependence of the 
QCD coupling (see Eq.\,(\ref{alphas})). The $\mu_Q$ dependence is canceled, order-by-order, by the $\mu_Q$ dependence in the hard factor
$H(\alpha_S)$\,\footnote{In case of the application of the unitarity constraint, the cancellation of the $\mu_Q$ dependence  involves also the
finite component of the distribution.}.
\item
The functional form of the functions $g_n(\lambda)$,
$n=1,2,3,4$, in 
Eqs.\,(\ref{g1}, \ref{g2}, \ref{g3}, \ref{g4}) is exactly the same as the one of the corresponding functions in transverse-momentum resummation in hadronic collisions\,\cite{Collins:1981uk,Kodaira:1981nh,Bozzi:2005wk} (see e.g.\ Refs.\,\cite{Bozzi:2005wk,Camarda:2023dqn}).
Note however that the $B_i$ coefficients are different
in the two cases beyond NLL, i.e. for $i>1$.
We also believe that, at very high orders, 
the equality of the $A_i$
coefficients in the two processes
can be violated%
\footnote{
A deeper criticism to standard
EEC resummation also applies.
The basic resummation rule
$\alpha_S \mapsto \alpha_S(k_\perp)$
is strictly valid only for
{\it inclusive gluon-decay quantities},
in which one does not distinguish 
between partons coming from (timelike)
gluon splitting, such as 
$g^*\mapsto gg, q \bar{q},\cdots$.
According to EEC definition, all
particle pairs have to be considered,
in particular also those involving
a generic parton and a parton
from gluon splitting
(these pairs have also been neglected in
the renormalon calculation in \cite{Schindler:2023cww}).
}.
The physical reason for the above similarity is the following.
At small angles, the transverse momentum 
of a parton is given by:
\beq
k_\perp \, \sim \, E \, \theta,
\eeq
where $E$ is the particle energy.
The definition of the EEC function 
involves a weighting factor $E/Q$,
implying that small particle energies
are suppressed.
The transverse momenta which dominate
the EEC are then 
\beq
k_\perp \, \sim \, Q \, \theta.
\eeq
Therefore angular variables in the EEC
naturally replace particle transverse 
momenta.

\end{enumerate}
The evaluation of the functions $g_n(\lambda)$ up to $n=4$ included,
requires the knowledge of 
the functions $A(\alpha_S)$ and $B(\alpha_S)$ on the  r.h.s.\ 
of Eqs.(\ref{eqA},\ref{eqB}) up to the $A_4$ and  $B_3$ coefficients respectively, together
with the coefficients of the QCD $\beta$-function 
up to $\beta_3$ (see Eq.(\ref{beta})).
These coefficients will be given in Sec.\,\ref{sec_resum_coef}. 


\subsection{Perturbative expansion of the resummed term}

The resummed part of the distribution in Eq.(\ref{EEC-b})
expanded in powers of $\alpha_S$ up to
third order produces the following fixed-order (f.o.) expansion:
\bea
\label{EEC-exp}
\frac{1}{\sigma_{\rm tot}} \,\frac{ d \Sigma_{\rm (res.)}}{dz}\Bigg|_{\rm f.o.}
\!\!&=& \frac{1}{2} \, \delta(1 \, - \, z)  
\, + \, \mathcal{A_{\rm (res.)}}(z) \, \frac{\alpha_S}{\pi} 
\, + \, \mathcal{B_{\rm (res.)}}(z)  \left( \frac{\alpha_S}{\pi} \right)^2 \nonumber\\
& + & \mathcal{C_{\rm (res.)}}(z)  \left( \frac{\alpha_S}{\pi} \right)^3 
\, + \, \mathcal{O}\left(\alpha_S^4\right)\,;
\eea
where:
\bea
\label{EEC-expA}
\mathcal{A_{\rm (res.)}}(z) &=&
-  \frac{A_1}{4}  I_2(z)
 -  \frac{{B}_1}{2}  I_1(z)  +  \frac{H_1}2  \delta(1-z);
\\
\label{EEC-expB}
\mathcal{B_{\rm (res.)}}(z) &=&  \frac{A_1^2}{16}  I_4(z)  +   
\frac{A_1}{2} \bigg(\frac{{B}_1}2  -  \frac{\beta_0}3 \bigg) I_3(z)
- \frac14 \bigg[
  A_2 - {B}_1^{2} + {B}_1 \beta_0 + A_1 H_1 
  \bigg] I_2(z) 
\nn\\
&-&  \frac{1}{2}\left[
{B}_2 + {B}_1 
H_1 
\right] I_1(z)  +  \frac{H_2}2  \delta(1-z);
\\
\label{EEC-expC}
\mathcal{C_{\rm (res.)}}(z) &=&
 - \frac{A_1^3}{96}  I_6(z)
-  \frac{A_1^2}{4}  
\left(\frac{{B}_1}4 - \frac{\beta_0}3\right) I_5(z) 
\nn\\
&+& \frac{A_1}8 \left[
A_2 - {B}_1^2 + \frac{7 {B}_1 \beta_0}3 - \beta_0^2 
+ A_1 \frac{H_1}2 
\right] I_4(z) 
\nn\\
&+& 
\left[  \frac{A_2 {B}_1}2 + \frac{A_1 {B}_2}2- \frac{{B}_1^3}6 - \frac{A_1\beta_1}3 - \frac23 A_2 \beta_0 
  +\bigg(\frac {{B}_1}2 
- \frac{\beta_0}3\bigg) \bigg(A_1H_1 + {B}_1 \beta_0 
\bigg)    
\right] 
\frac{I_3(z)}{2} \!\!\!\!\!\!\!\!\!\!
\nn\\
&+& 
\left[
  -\frac{A_3}2 + ({B}_1 - \beta_0)\bigg(\frac{{B}_1 H_1}2 +{B}_2 
  \bigg) - \frac{{B}_1 \beta_1}2 - \frac{A_2 H_1}2 
 - \frac{A_1 H_2}2 
\right] \frac{I_2(z)}{2}  
\nn\\
& -&   \left(
    {B}_3 + {B}_2 H_1 + {B}_1 H_2
\right)
\frac{I_1(z)}{2} 
+\frac{H_3}2  \delta(1-z)\,,
\eea
where we have set $\mu_R=\mu_Q=Q$. The complete dependence of the functions $\mathcal{A_{\rm (res.)}}(z)$, $\mathcal{B_{\rm (res.)}}(z)$ and $\mathcal{C_{\rm (res.)}}(z)$
on the renormalization and resummation scales can be straightforward obtained by expanding Eq.\,(\ref{EEC-b}) and using the results in Eqs.\,(\ref{sud2}-\ref{g4}).
The functions $I_n(z)$ are the Fourier-Bessel transform of the logarithmic terms in $b$-space:%
\footnote{
Note that these improper integrals are divergent, because the integrands
are not infinitesimal at infinity (they rather diverge). 
A solution to this problem involves
considering the functions $I_n(z)$ {\it distributions},
i.e.\ {\it generalized functions}.
}	
\beq
\label{Ifun}
I_n(z) \, \equiv \,
\int_0^\infty d(Qb) \frac{Qb}2\,J_0\left(\sqrt{1-z} \, Q b\right)
\,\ln^n\left(\frac{Q^2 \, b^2}{b_0^2} \right);
\qquad n = 0, 1, 2, \cdots \,.
\eeq
The explicit evaluation of these integrals gives\,\cite{Bozzi:2005wk}:
\bea
\label{Ifunn}
I_0(z) &=& \delta(1-z), \qquad
I_1(z)=-\left[\frac1{1-z}\right]_+ \,, \qquad
I_2(z)=2\left[\frac{\ln(1-z)}{1-z}\right]_+\,,\quad
\nn\\
I_3(z)&=&-4z_3\delta(1-z)-3\left[\frac{\ln^2(1-z)}{1-z}\right]_+ \,, 
\qquad
I_4(z)=4\left[\frac{\ln^3(1-z)+4z_3}{1-z}\right]_+\,,\quad
\nn\\
I_5(z)&=&-48z_5\delta(1-z)-5\left[\frac{\ln^4(1-z)+16z_3\ln(1-z)}{1-z}\right]_+\,,
\nn\\
I_6(z)&=&160z_3^2\delta(1-z)+6\left[\frac{\ln^5(1-z)+40z_3 \ln^2(1-z)+48z_5}{1-z}\right]_+\,.
\eea
The application of the unitarity constraint $L\mapsto \widetilde L$ of Eq.\,(\ref{Ltilde}) modifies the
asymptotic expansion of Eq.\,(\ref{EEC-exp}) in the following way
\bea
\label{EEC-exptil}
\frac{1}{\sigma_{\rm tot}} \,\frac{ d \widetilde{\Sigma}_{\rm (res.)}}{dz}\Bigg|_{\rm f.o.}
\!\!&=& \frac{1}{2} \, \delta(1 \, - \, z)  
\, + \, \mathcal{\widetilde{A}_{\rm (res.)}}(z) \, \frac{\alpha_S}{\pi} 
\, + \, \mathcal{\widetilde{B}_{\rm (res.)}}(z)  \left( \frac{\alpha_S}{\pi} \right)^2 \nonumber\\
& + & \mathcal{\widetilde{C}_{\rm (res.)}}(z)  \left( \frac{\alpha_S}{\pi} \right)^3 
\, + \, \mathcal{O}\left(\alpha_S^4\right)\,;
\eea
where the functions $\mathcal{\widetilde{A}_{\rm (res.)}}(z)$,
$\mathcal{\widetilde{B}_{\rm (res.)}}(z)$ and $\mathcal{\widetilde{C}_{\rm (res.)}}(z)$
can be obtained from the corresponding ones in Eqs.\,(\ref{EEC-expA}--\ref{EEC-expC})  simply replacing $I_n(z)$
with the following functions: 
\beq
\label{Ifuntilde}
\widetilde{I}_n(z) \, \equiv \,
\int_0^\infty d(Qb) \frac{Qb}2\,J_0\left(\sqrt{1-z} \, Q b\right)
\,\ln^n\left(\frac{Q^2 \, b^2}{b_0^2} +1\right);
\qquad n = 0, 1, 2, \cdots \,.
\eeq
The explicit evaluation of the functions $\widetilde{I}_n(z)$ is more involved than that of $I_n(z)$. However they can
be expressed in terms of
the modified Bessel function of imaginary argument $K_\nu(z)$ (see Appendix B of Ref.\,\cite{Bozzi:2005wk}). We observe that
the functions $I_n(z)$ and $\widetilde{I}_n(z)$ have the same large-$z$ behavior while they differ for small $z$.

We can now compare the expressions on the r.h.s.\ of Eq.(\ref{EEC-exp}) with 
(explicit) fixed-order calculations.
By comparing the first-order term $\mathcal{A_{\rm (res.)}}(z)$
with the large-$z$ contributions of the function $\mathcal{A}(z)$ 
on the r.h.s.\ of Eq.(\ref{eq_Afunc_complete}),
one directly determines the first-order coefficients
$A_1$, $B_1$ and $H_1$.

The comparison of the second-order function $\mathcal{B_{\rm (res.)}}(z)$
with the large-$z$ part of the analytic computation
of $\mathcal{B}(z)$ made in \cite{Ebert:2020sfi,Dixon:2018qgp},
has two different aspects:
\begin{enumerate}
\item
an {\it explicit and non-trivial (passed) check} 
of the coefficients of the integrals
$I_4,I_3,I_2,I_1$, by using the resummation coefficients $A_2$\,\cite{Kodaira:1982cr} and $B_2$\,\cite{deFlorian:2004mp}
already known from the literature;
\item
the {\it extraction} of the coefficient $H_2$
(note that a contribution to the $\delta(1-z)$ coefficient also
comes from $I_3$).
\end{enumerate} 
The comparison of the third-order function $\mathcal{C_{\rm (res.)}}(z)$
with the large-$z$ analytic computation of $\mathcal{C}(z)$
made in SCET in \cite{Ebert:2020sfi},
is similar to the previous one; after a (highly non-trivial) check
of the resummation formula,
passed also in this case,
we extracted the coefficients $B_3$ and $H_3$.
We also checked the consistency,
in the back-to-back region,
of the analytic $\mathcal{O}(\alpha_S^3)$
SCET computation 
against the corresponding numerical computation
in full QCD.
A specific check has also been made.
By replacing 
the third-order coefficient
$A_3$ in the resummation formula
 with the corresponding third-order
coefficient appearing in threshold resummation (i.e.\ the three-loop cusp anomalous dimension $\Gamma_2$),
expanding to $\mathcal{O}(\alpha_S^3)$ and comparing again with the numerical QCD computation\,\cite{Tulipant:2017ybb},
we find that the $\chi^2$ increases by over a factor of six
with respect to the previous case.
That is a numerical evidence, inside full QCD, of the correctness
of the SCET relation between the two above quantities \cite{Becher:2010tm}:
\beq
\label{eq_A3qtand A3th}
A_3 \, = \, \Gamma_2
\, + \, \frac{1}{8} \beta_0 \, d_2;
\eeq
where:
\beq
d_2 \, = \, C_F\left[
C_A \left( \frac{808}{27} - 28 \, z_3  \right)
\, - \, \frac{112}{27}\, n_f
\right ] .
\eeq


\section{Resummation coefficients}
\label{sec_resum_coef}

The explicit values of the coefficients 
$A_1$, $A_2$, $A_3$, $A_4$, $B_1$, $B_2$, $B_3$, $H_1$, $H_2$ and $H_3$
needed up to N$^3$LL+NNLO accuracy are:
\bea
A_1 &=& C_F\,,
\\
A_2 &=& C_F
\left[
C_A \left( \frac{67}{36} - \frac{z_2}{2} \right)
- \frac{5}{18} n_f
\right]\,,
\\
A_3 &=& C_F
\Bigg[
C_A^2 \left(
\frac{15503}{2592} - \frac{67}{36} z_2 
+ \frac{11}{8} z_4 - \frac{11}{4} z_3
\right)
+ C_A n_f \left(
- \frac{2051}{1296}
+ \frac{5}{18} z_2
\right) 
\nonumber\\
&+&  C_F n_f \left( 
- \frac{55}{96} + \frac{z_3}{2}
\right)
+ \frac{25}{324} n_f^2
\Bigg]\,,
\\
A_4 &=& C_F
\Bigg[
C_A^3 
\left(
- \frac{z_3^2}{16} + \frac{33}{16} z_2 z_3
- \frac{16475}{864} z_3
- \frac{6613}{864}  z_2
+ \frac{1859}{384} z_4 + \frac{1925}{288}  z_5 - \frac{313}{96}  z_6 \right.\nn\\
&+& \left.  \frac{4520317}{186624}
\right) 
+C_A^2 n_f
   \left(\frac{z_2 z_3}{8} + \frac{1757}{864}  z_2
   + \frac{689}{288} z_3   
   - \frac{11}{96}  z_4
- \frac{85}{144}  z_5 - \frac{571387}{62208}\right) 
\nonumber\\
&+&  C_A C_F  n_f \left(
 - \frac{1}{2} z_2  z_3 + \frac{55}{96} z_2
 + \frac{7}{2} z_3
 + \frac{5}{8} z_5 - \frac{100225}{20736}\right) 
\nonumber\\
&+&  C_A  n_f^2
\left( - \frac{47}{432} z_2
+ \frac{31}{144} z_3
   - \frac{3}{32} z_4 + \frac{58045}{62208}\right) 
\nonumber\\
&+&  C_F^2 n_f
   \left(\frac{37}{48}  z_3 - \frac{5}{4} z_5 + \frac{143}{576}\right) 
+ C_F  n_f^2
   \left(\frac{7001}{10368} - \frac{13}{24} z_3 \right)
   - \left( \frac{z_3}{54}  
   + \frac{125}{5832} \right)  n_f^3
\Bigg] 
\nonumber\\
&+& 
\left(z_2 - \frac{z_3}{3} - \frac{5}{3} z_5 \right) n_f
   \frac{d_{FF}^{(4)} }{N_c}  
+ \left( - \frac{3}{2} z_3^2 + \frac{z_3}{6}
- \frac{z_2}{2} + \frac{55 }{12} z_5 - \frac{31 }{8} z_6
\right)
   \frac{d_{FA}^{(4)} }{N_c}\,.
\eea
\bea
\label{b1b2b3}
B_1 &=& -  \frac{3}{2}  C_F \,,\\
B_2 &=& C_F
\Bigg[
C_A \left(
- \frac{35}{16} + \frac{11}{4} z_2 + \frac{3}{2} z_3
\right)
+ C_F\left( - \frac{3}{16} + \frac{3}{2} z_2 - 3 z_3  \right)
+ n_f \left( \frac{3}{8} - \frac{z_2}{2}\right)
\Bigg], 
\\
B_3 &=& C_F
\Bigg[
 C_A n_f
 \left(- \frac{377}{72} z_2
- \frac{7}{12}  z_3
+ \frac{7}{12} z_4 + \frac{707}{432} \right) 
\nn\\
&+& C_A C_F \left( - \frac{1}{2} z_2  z_3 
+ \frac{161}{48} z_2 + \frac{49}{6} z_3 - \frac{413}{48} z_4
- \frac{15}{4} z_5 - \frac{97}{144}\right)  
\nonumber\\
&+& C_A^2 \left(
\frac{1105}{72} z_2
+ \frac{17}{12} z_3
- \frac{163}{48} z_4
- \frac{5}{4} z_5 - \frac{4241}{864}
\right)  
\nonumber\\
&+&  C_F n_f
\left( - \frac{2}{3} z_2 - \frac{5}{3} z_3
+ \frac{31}{24} z_4 + \frac{505}{576} \right)
+ n_f^2 \left(
\frac{7}{18} z_2
+ \frac{z_3}{6}
- \frac{49}{432} \right) 
\nonumber\\
&+&  C_F^2
\left(
z_2  z_3 - \frac{9}{16} z_2 - \frac{17}{8} z_3
- \frac{9}{2} z_4 + \frac{15}{2} z_5 - \frac{29}{64}\right)
\Bigg] \,.
\eea
\bea
H_1 &=& -  C_F \left( \frac{11}{4}  +  z_2 \right)%
\,,
\\
H_2 &=& C_F \Bigg[
C_A \left(
- \frac{2635}{288} - \frac{13}{9} z_2 - z_4
+ \frac{82}{9} z_3
\right)  +
C_F \left(
\frac{371}{96} + \frac{55}{8} z_2 + 6 z_4 - 13 z_3
\right) 
\nonumber\\
&+&  n_f \left(
\frac{215}{144} + \frac{z_2}{9}
- \frac{z_3}{9}
\right)
\Bigg]%
\,,
\\
H_3 &=& 
C_F \Bigg[
C_A^2 \left(
- \frac{437 }{72} z_2 z_3
- \frac{701 }{2592} z_2
- \frac{7 }{4} z_3^2
+ \frac{157393}{2592}  z_3
+ \frac{3815 }{576} z_4
- \frac{625 }{24} z_5
+ \frac{59}{32}  z_6 \right. \nn\\
&-& \left. \frac{493351}{10368}
\right)
+ C_A n_f 
\left(
  \frac{73}{36}  z_2  z_3
- \frac{2257 }{1296} z_2
- \frac{6113}{648}  z_3
- \frac{145 }{144} z_4
+ \frac{25}{12}  z_5
+ \frac{19453}{1296}
\right) 
\nonumber\\
&+& C_A C_F 
\left(
- \frac{127 }{36} z_2  z_3 
+ \frac{1601 }{48} z_2
+ 5 z_3^2
- \frac{4103}{48} z_3 %
+ \frac{299 }{36} z_4
+ \frac{121}{12}  z_5
+\frac{11}{16}  z_6
+ \frac{26945}{864}
\right) 
\nonumber\\
&+& C_F^2
   \left(
22 z_2  z_3
- \frac{1553}{96}  z_2
- \frac{14}{3}  z_3^2
+ \frac{1111}{48}  z_3
- \frac{385}{8}  z_4
+ \frac{125}{3}  z_5
- 21 z_6
- \frac{2353}{384}
\right) 
\nonumber\\
&+& C_F n_f
   \left(
   - \frac{55 }{18} z_2  z_3 
- \frac{1207}{288}  z_2
+ \frac{809}{72}  z_3 %
+ \frac{29 }{144} z_4
- \frac{8 }{3} z_5
- \frac{10213}{3456}
\right) 
\nonumber\\
&+&  n_f^2
   \left(
  \frac{181 }{648} z_2
- \frac{25}{324}z_3 
- \frac{z_4}{18} 
- \frac{715}{648}
\right) 
\Bigg] \nn\\
&+& \frac{ d_{FF}^{(3)} N_{FV}}{N_c}
 \left(
\frac{19     }{96}  z_3
 - \frac{ 5   }{12}  z_5
+ \frac{1}{192}
+ \frac{5}{32} z_2
- \frac{z_4}{64}
\right)\,. 
\eea
The standard color factors of $SU(N_c)$ are 
$C_F \, = \, (N_c^2 \, - \,1)/(2 N_c) \, = \, 4/3$,
$C_A \, = \, N_c \, = \, 3$ for $N_c=3$ colors and $n_f$ is the number of $QCD$ active
(effective massless) flavors at the scale $Q$ ($n_f=5$ for $Q=m_Z$).  
The coefficients $z_n$  are the values of the Riemann zeta-function at the integer points $n=2,3,4,5,6$:
$z_2=\pi^2/6=1.64493\cdots$, $z_3=1.20206\cdots$, $z_4=\pi^4/90=1.08232\cdots$, $z_5=1.03693\cdots$ and $z_6=\pi^6/945=1.01734\cdots$.
$n_f$ is the number of QCD active
(i.e. effective massless) flavors at the scale $Q$ 
($n_f=5$ for $Q=m_Z$). 

The following color factors are present in the expression of the coefficients $H_3$ and $A_4$ respectively:
\bea
d_{FF}^{(3)} &\equiv& d^{abc}_F d^{abc}_F = \frac{ (N_c^2 - 4)(N_c^2 - 1)}{N_c}, 
\nn\\
d_{FF}^{(4)} &\equiv& d^{abce}_F d^{abce}_F = \frac{\left( N_c^4 - 6 N_c^2 + 18 \right)  \left( N_c^2 - 1 \right) }{ 96 N_c^2 };
\nn\\
d_{FA}^{(4)} &\equiv& d^{abce}_F d^{abce}_A = \frac{ N_c \left( N_c^2 + 6 \right)  
\left( N_c^2 - 1 \right) }{ 48 }.
\eea
where $d^{abc}_R$ and $d^{abce}_R$ are the third-order and fourth-order
totally symmetric
tensors of the fundamental ($R=F$) and adjoint ($R=A$)
representation of $SU(N_c)$\,\cite{vanRitbergen:1998pn}.
The factor $N_{FV}$ originates from diagrams where the virtual gauge boson does not couple
directly to the final-state quarks but to a closed quark loop. It is thus proportional to the charge weighted sum of the
quark flavors. For (single) photon exchange 
with $n_f=5$,
its values is simply
\beq
N_{F\gamma} \, = 
\frac{\sum_{f=1}^{5} e_f}{e_q}
\, = \, \frac{1}{3 \, e_q},
\eeq
where $e_q$ is the electric charge of the final-state quark in the Born-level cross section.
Its exact value in case of $\gamma^*/Z$ exchange turns out to be
irrelevant for phenomenological applications (see comments below Eq.\,(\ref{h2num})).

Finally the coefficients $\beta_0$, $\beta_1$, $\beta_2$\,\cite{Tarasov:1980au,Larin:1993tp}
and $\beta_3$\,\cite{vanRitbergen:1997va,Czakon:2004bu}
 of the QCD $\beta(\alpha_S)$ function are also needed. In our conventions they
are explicitly given by:
\bea
\beta_0 &=& \frac{11 C_A \, - \, 2 n_f}{12}\,, 
\nn\\
\beta_1 &=& \frac{1}{24} \left( 17 C_A^2 - 5 C_A n_f - 3 C_F n_f \right)\,, 
\\
\beta_2 &=&  \frac{1}{64} 
\left[
\frac{2857}{54} \, C_A^3
\, - \, 
\left(
\frac{1415}{54} C_A^2 \, + \, \frac{205}{18} C_A C_F \, - \, C_F^2
\right) n_f
\, + \,
\left(
\frac{79}{54} C_A + \frac{11}{9} C_F 
\right) n_f^2
\right]\,,
\nonumber\\
\beta_3 &=&
\frac{1093}{186624}n_f^3 + n_f^2 \left(\frac{809 z_3}{2592}
+ \frac{50065}{41472}\right) + n_f
   \left( - \frac{1627 z_3}{1728} - \frac{1078361}{41472} \right)
+\frac{891}{64} z_3 + \frac{149753}{1536}\,,
\nonumber
\eea
where, for simplicity's sake, in the $\beta_3$ coefficient we
have replaced the explicit value of the color factors.



The coefficient $A_1$ is the leading one in the resummation formula.
Being positive, it is responsible for the well-known Sudakov suppression
of elastic (non-radiative) channels
and it was already included in the original papers
on the EEC resummation\,\cite{Basham:1978bw}. 
Unlike the higher-order coefficients, it is {\it kinematical},
i.e.\ it does not involve "irreducible" many soft-gluon effects.
The coefficient $A_2$ has been originally computed in \cite{Kodaira:1982cr},
by looking at the soft-enhanced part of the next-to-leading splitting
function $P_{qq}^{(1)}(z)$.
The third-order and fourth-order coefficients $A_3$ and $A_4$ can be
derived for $q_T$-resummation from the results of  
Refs.\,\cite{Moch:2004pa,Becher:2010tm,Moch:2018wjh,vonManteuffel:2020vjv,Li:2016ctv}.

The coefficient $B_1$, like $A_1$, is kinematical and has been 
already included in the original resummation papers.
Being negative, it has an opposite effect to the coefficient $A_1$, i.e.\ it reduces
the Sudakov suppression. 
The coefficient $B_2$ has been originally computed 
around twenty years ago in \cite{deFlorian:2004mp}
by using approximate QCD matrix elements
having the correct relevant infrared limits. 
The expression for the third-order coefficient $B_3$ 
on the r.h.s.\ of Eq.(\ref{b1b2b3}), within the resummation formalism in full QCD, is new.

Let us remark that, in order to reach full N$^3$LL  accuracy for the physical cross-section (and not merely in the exponent of the Sudakov form factor), the hard-virtual coefficient function $H(\alpha_S)$ has to be evaluated up to third order included.
At NLL, the coefficient $H_1$ is required because
the multiplication of the term 
$\alpha_S H_1$ with, for example, the term $\alpha_S L^2$ in $\exp{[L g_1(\lambda)]}$
(see Eq.\,(\ref{sud2})) generates
terms of the same order of the NLL corrections in the function $g_2(\lambda)$. 
For the same reason, the coefficients $H_2$ and $H_3$ are  needed 
at NNLL and N$^3$LL accuracy respectively.

The coefficient $H_1$ was known since long time\,\cite{Basham:1978bw}, 
while the expressions of the coefficients $H_2$ and $H_3$ 
are, within the QCD resummation formalism, 
as far as we know, new.  

The higher-order $H_2$ and $H_3$ coefficients explicitly depend 
both on the resummation and
renormalization scales, while $H_1$ only depends on $\mu_Q$.
The explicit renormalization-scale dependence is exactly canceled,
order-by-order in perturbation theory, by the renormalization scale
dependence of the QCD coupling (see Eq.\,(\ref{alphas})).
The resummation-scale dependence of $H(\alpha_S)$ is canceled by
the $\mu_Q$ dependence
of the Sudakov form factor $S(Q,b)$ and (in case of unitarity constraint)
of the finite component of the distribution,
leaving a residual subleading logarithmic dependence on $\mu_Q$.
In Sec.\,\ref{sec_scales} we will exploit the $\mu_R$ and $\mu_Q$
dependence in order to estimate the perturbative uncertainty
of our calculation due respectively to the renormalization and resummation procedures.


\subsection{Numerical values of the resummation coefficients}

We now comment on the numerical values of the resummation coefficients in the case of $n_f=5$ active flavors.

The coefficients $A_1$ and $A_2$ are 
positive:
\beq
\frac{A_1}{\pi} \, \simeq \, 0.42;
\quad
\frac{A_2}{\pi^2} \, \simeq \, 0.23;
\qquad (n_f = 5).
\eeq
On the contrary, coefficients  $A_3$ and $A_4$
turns out to be negative:
\beq
\frac{A_3}{\pi^3} \, \simeq \, - \, 0.21;
\quad
\frac{A_4}{\pi^4} \, \simeq \, - \, 0.76  \qquad (n_f=5).
\eeq
Therefore, $A_1$S and $A_2$ tend to suppress the rate
close to Born kinematics 
(a back-to-back quark-antiquark pair, 
$\theta_{q\bar{q}} = \pi$, in the center-of-mass frame),
producing the well-known Sudakov suppression,
while the coefficients $A_3$ and $A_4$ tend to (slightly) enhance it.

As far as the single-log function $B(\alpha_S)$
is concerned, we find that $B_1$ is negative,
\beq
\frac{B_1}{\pi} \, \simeq \, - \, 0.64,
\eeq
while the second and third-order coefficients
are both positive and larger in size:
\beq
\frac{B_2}{\pi^2} \, \simeq  \, 1.14;
\qquad 
\frac{B_3}{\pi^3} \, \simeq  \, 2.76
\qquad (n_f=5).
\eeq
Therefore, unlike $B_1$, the higher-order coefficients $B_2$ and $B_3$ tend 
to suppress the rate close to Born kinematics.
Compared to $B_1$, the coefficient $B_2$ is of the expected size, as:
\beq
\left| \frac{B_2}{\pi B_1} \right| \, \simeq \, 1.79 
\quad \mathrm{and} \quad \frac{C_A}{C_F} \, = \, 2.25.
\eeq
Note also that:
\beq
\frac{B_3}{\pi B_2} \, = \, 2.43,
\eeq
close to the previous ratio.

Let's finally consider the hard coefficient function $H(\alpha_S)$.
The first-order term is negative and relatively  large in size:
\beq
\frac{H_1}{\pi} \, \simeq \, - \, 1.87.
\eeq
With $\alpha_S(m_Z)=0.118$, for example, the first-order correction reduces
the lowest-order coefficient by over $20\%$:
\beq
1 \, + \, H_1 \frac{\alpha_S}{\pi} \, \simeq \, 0.78.
\eeq
The second-order coefficient $H_2$ is positive and it is of similar
size to $H_1$:
\beq
\label{h2num}
\frac{H_2}{\pi^2} \, \simeq \, 1.46
\qquad (n_f = 5).
\eeq
The coefficient $H_3$ is negative and smaller in size.
In the case of photon exchange, for example, 
and for a down-type quark:
\beq
\frac{H_3}{\pi^3} \, \simeq \, - \, 0.62
\qquad (n_f = 5).
\eeq
In particular the impact of the part proportional to $N_{FV}$ is, for $\alpha_S(m_Z)=0.118$,
\beq
\left(\frac{\alpha_S}{\pi}\right)^3\, H_3\bigg|_{N_{FV}} \, \simeq \, - \, 1.2\, \times\,10^{-5}\, N_{FV}\,.
\eeq
Therefore the effect of the diagrams where the virtual gauge boson does not couple
directly to the final state quarks  is completely negligible.
  

\section{The finite (remainder) component}
\label{sec_remainder}

The finite component, or remainder function, of the EEC distribution,
appearing on the r.h.s.\ of Eq.\,(\ref{EEC-dec}),
is a short-distance, process-dependent function, 
which dominates the cross section away from the back-to-back region.
Therefore its knowledge is necessary in order to reach a uniform theoretical 
accuracy over the entire kinematical range.
The remainder function can be obtained from
the fixed-order expression in Eq.(\ref{EECfot}), by subtracting the
perturbative expansion
of the resummed component, truncated at the same order:
\beq
\label{EEC-asy}
\frac{1}{\sigma_{\rm tot}} \, \frac{ d \Sigma_{\rm (fin.)} }{dz} \, = \,
\frac{1}{\sigma_{\rm tot}} \, \frac{ d \Sigma }{dz}
\, - \,
\frac{1}{\sigma_{\rm tot}} \, \frac{ d \Sigma_{\rm (res.)} }{dz}\Bigg|_{\rm f.o.}\,,
\eeq
where the subscript ``f.o.'' indicates the customary fixed-order expansion.
In order to ensure that the finite component is free from
large infrared logarithms, we {\it require} that the second term in the r.h.s.\ of
Eq.(\ref{EEC-asy}) exactly contains {\it all the
infrared logarithmic corrections} (see Eq.(\ref{logs})) of the 
fixed-order expansion of the EEC distribution. 
In principle, this {\itshape matching requirement}
can always be fulfilled if the logarithmic accuracy of the resummed term
is sufficiently high. 
In particular, the remainder functions at LO ($\mathcal{O}(\alpha_S)$), NLO
($\mathcal{O}(\alpha_S^2)$) and NNLO ($\mathcal{O}(\alpha_S^3)$)
have to be matched with the NLL, NNLL and N$^3$LL  resummed terms respectively. 
We thus refer to NLL+LO, N$^2$LL+NLO and N$^3$LL+NNLO perturbative accuracies.

The remainder function, for $z>0$, has a perturbative expansion
of the form:
\footnote{
If we consider also the lower endpoint $z=0$, then the remainder function
also contains a zero-order term $\left(\mathcal{O}(\alpha_S^0)\right)$ 
given by $1/2 \, \delta(z)$ (cfr. Eq.(\ref{EECfot})),
as well as higher-order corrections of the same form,
together with plus regularizations of the non-integrable
terms for $z \to 0^+$ (i.e.\ the terms
$\propto 1/z$, cfr. Eq.\,(\ref{eq_Afunc_complete}))%
.}
\bea
\label{EEC-fin}
\frac{1}{\sigma_{\rm tot}} \,\frac{ d \Sigma_{\rm (fin.)}}{dz}
&=& 
\mathcal{A_{\rm (fin.)}}(z) \, \frac{\alpha_S}{\pi} 
 +  \mathcal{B_{\rm (fin.)}}(z)  \left( \frac{\alpha_S}{\pi} \right)^2
 +  \mathcal{C_{\rm (fin.)}}(z)  \left( \frac{\alpha_S}{\pi} \right)^3
 +   \mathcal{O}\left(\alpha_S^4\right)\nn\\
 &=& 
 \left[\mathcal{A}(z)-\mathcal{A_{\rm (res.)}}(z)\right]  \frac{\alpha_S}{\pi} 
 +  \left[\mathcal{B}(z)-\mathcal{B_{\rm (res.)}}(z)\right]  \left( \frac{\alpha_S}{\pi} \right)^2\nn\\
 &+&  \left[\mathcal{C}(z)-\mathcal{C_{\rm (res.)}}(z)\right]  \left( \frac{\alpha_S}{\pi} \right)^3
 +   \mathcal{O}\left(\alpha_S^4\right)\,.
\eea


\subsection{Leading Order}

The LO term of the remainder is derived from Eq.(\ref{eq_Afunc_complete});
it has, for $z>0$, this simple analytic expression:%
\beq
\label{EEC-fin2}
\mathcal{A_{\rm (fin.)}}(z) =  
-  \frac{2}{3 \, z^5} \left( z^4+z^3-3 z^2+15 z-9 \right) \ln (1-z)
 -  \frac{z^3+z^2+7 z-6}{ z^4}
\qquad (0<z<1),
\eeq
where, for simplicity's sake, we have replaced the explicit value of $C_F=4/3$.


\subsection{Next-to-Leading Order}

The NLO remainder function is extracted from the analytic computation in 
\cite{Dixon:2018qgp}.
In order to obtain a more compact formula,
we substitute in $\mathcal{B_{\rm (fin.)}}(z)$ 
the explicit values of the color factors
(the general formulae can be found in \cite{Dixon:2018qgp}).
The expression for $z>0$ at the central value of the renormalization scale $\mu_R=Q$ reads:
\bea
\mathcal{B_{\rm (fin.)}}(z) &=&
 \frac{1080 z^6-3240 z^5+4164 z^4-2924 z^3+1134
   z^2-229 z+1 }{9 z(1-z)} \, z_3 \, +
\nonumber\\
&+&\frac{1 }{54 (1-z) z^5} 
\big(
+ 1080 z^{10}-3240 z^9+4164 z^8-2934 z^7+1155
   z^6 +
\nonumber\\
   && \qquad \qquad\qquad\,\, -251 z^5+8 z^4-33 z^3+153 z^2-177 z+41
\big)
\ln^3(1-z) +
\nonumber\\
&+& \frac{161 z^6-274 z^5+2020
   z^4+1505 z^3+4165 z^2-5176 z+3404 }{540 z^5} \ln^2(1-z) +
\nonumber\\
&-& \frac{360 z^6-1080 z^5+1388
   z^4-980 z^3+386 z^2-78 z+3 }{18 (1-z) z} \ln(z) \ln^2(1-z) +
\nonumber\\
&-& \frac{1 }{3240 z^5} 
\big(
+ 129600 z^8-226800
   z^7+190080 z^6-72672 z^5+5936 z^4 +
\nonumber\\   
&& \qquad \qquad  - 9878 z^3+38043
   z^2-117752 z+68243
\big)
\ln(1-z) +
\nonumber\\
&+& \frac{1}{54 (1-z) z^4} 
\big( + 360 z^9-1080 z^8+1388 z^7-980 z^6+380 z^5-93
   z^4+z^3 +
\nn\\   
&& \qquad \qquad\qquad
-2 z^2+6 z-6 \big) \pi^2 \ln (1-z) +
\nonumber\\
&+&\frac{1}{9 (1-z) z^5} 
\big(
+ 360 z^{10}-1080 z^9+1388 z^8-974 z^7+383 z^6-71 z^5+2 z^4 +
\nn\\
&& \qquad\qquad\qquad -33 z^3+153 z^2-177 z+41 \big) 
\text{Li}_2(z) \ln (1-z) +
\nonumber\\
&+& \pi^2 \frac{ 1 }{3240 (1-z) z^5}
\big( 
-322 z^7+630 z^6-5038 z^5+650 z^4+4954 z^3 +
\nn\\
&& \qquad\qquad\qquad\qquad - 22850 z^2+30827 z-12391 \big) +
\nonumber\\
&+&\frac{1}{6480 (1-z) z^4}
\big(
+ 64800 z^7-129600 z^6+115836 z^5-43548 z^4+6460 z^3 +
\nn\\
&& \qquad\qquad\qquad\quad\,\, + 143397 z^2-236547 z+79202 \big) +
\nonumber\\
&-&\frac{2
   z^5-z^4+2 z^3+z^2+3 }{18 z^4} \ln
   ^2\left(\frac{1+\sqrt{z}}{1-\sqrt{z}}\right) \ln
   \left(\frac{1-z}{z}\right) +
\nonumber\\
&+&\frac{ -360 z^5+1080 z^4-1388 z^3+974 z^2-377 z+69}{54 (1-z)} \pi ^2 \ln (z) +
\nonumber\\
&-&\frac{1}{3240 (1-z)
   z^4} 
\big(
+ 129600 z^8-291600 z^7+287280 z^6-146700 z^5 +
\nn\\   
&& \qquad\qquad\qquad
+ 35204 z^4-2386 z^3+67175 z^2-150039 z+74346
\big) \ln(z) +
\nonumber\\
&+&\frac{480 z^4-80 z^3+496 z^2+381 z+735 }{1080 z^{7/2}}
\ln \left(\frac{1-\sqrt{z}}{1+\sqrt{z}}\right) \ln(z) +
\nonumber\\
&+&\frac{1}{540 z^5} 
\big(
- 21600 z^8+32400 z^7-25680
   z^6+7680 z^5-1960 z^4 +
\nn\\   
&&\qquad\qquad   -720 z^3+4054 z^2-18436
   z+12391 \big) \text{Li}_2(1-z) +
\nonumber\\
&+&\frac{-480 z^4+80 z^3-496 z^2-381 z-735
   }{540 z^{7/2}} \text{Li}_2\left(-\sqrt{z}\right) +
\nonumber\\
&+&\frac{480 z^4-80 z^3+496 z^2+381
   z+735 }{540 z^{7/2}} \text{Li}_2\left(\sqrt{z}\right) +
\nonumber\\
&-&\frac{1}{540 (1-z) z^5} 
\big(
+ 21600 z^9-54000 z^8+58402 z^7-34230
   z^6+10688 z^5 +
\nn\\   
&& \qquad\qquad\qquad\quad  -2270 z^4+546 z^3+3808 z^2-13667
   z+5583
\big) \text{Li}_2(z) +
\nonumber\\
&-&\frac{360 z^6-1080 z^5+1388 z^4-980 z^3+386
   z^2-78 z+3 }{9 (1-z) z} \ln(z) \text{Li}_2(z) +
\nonumber\\
&+&\frac{4 \left(2 z^5-z^4+2 z^3+z^2+3\right)}{9 z^4} 
\text{Li}_3\left(-\frac{\sqrt{z}}{1-\sqrt{z}}\right)
 +
\nonumber\\
&+&\frac{4 \left(2 z^5-z^4+2 z^3+z^2+3\right)}{9 z^4} 
\text{Li}_3\left(\frac{\sqrt{z}}{\sqrt{z}+1}\right) +
\nonumber\\
&+&\frac{2(-6 z^7+3 z^6-7 z^5+z^4+33 z^3-153
   z^2+177 z-41) }{9 (1-z)
   z^5} \text{Li}_3(z) +
\nonumber\\
&-&\frac{1}{9 (1-z) z^5} 
\big(
1080 z^{10}-3240 z^9+4164 z^8-2940 z^7+1158 z^6-234 z^5 +
\nn\\
&& ~~ + \, 9 z^4-65 z^3+303 z^2-351 z+82
\big)
\text{Li}_3\left(-\frac{z}{1-z}\right)
\qquad (0<z<1),
\eea
where $\text{Li}_2(z)$ and $\text{Li}_3(z)$ denote the standard
dilogarithm and trilogarithm respectively.


\subsection{Next-to-Next-to-Leading Order}

Let's now discuss the (non  trivial) evaluation of the NNLO remainder function.
The latter is evaluated numerically for $0<z<1$ in the following way.
One first subtracts from the third-order (complete) QCD distribution $\mathcal{C}(z)$,
known as a numerical table,
the $\ln^k(1-z)/(1-z)$ terms, $k=0,1,\cdots,5$,
i.e.\ the large logarithms of $1-z$, which are all known analytically at $\mathcal{O}(\alpha_S^3)$ and are factorized, as already discussed, 
in the Sudakov form factor.
The next step is to {\it fit} the subtracted
$\mathcal{C}_{\rm (fin.)}(z)$ function.
The individual points of this (tabulated) function have indeed large statistical errors, being computed with a Monte Carlo numerical program. 
Since the statistical errors of the $\approx 100$ different points of the table can be assumed to be independent, 
by fitting these points with a reasonable function
without too many free parameters,
{\it a substantial reduction of the statistical fluctuations is expected
to occur.}
The functional form of the third-order remainder can be obtained by means
of the following considerations.
The first-order remainder behaves, for $z \to 1^-$, as $a_1 \ln(1-z)+a_0$ 
(with $a_1$ and $a_0$ constants of order one (or, if preferred, $\pi$)),
while the second-order remainder behaves, in the same limit, as 
$b_3\ln^3(1-z) + b_2 \ln^2(1-z) + b_1\ln(1-z) + b_0$
(with the $b_i$'s again of order one (or $\pi^2$)).
We {\it can conjecture} that, at order $n$, the remainder function
is a polynomial in $\ln(1-z)$ of order $2n-1$.
Indeed these terms all come from the $\left[\ln^k(1-z)/(1-z)\right]_+$ terms
contained in the Sudakov form factor, when multiplied by $1-z$.
A good fit for the third-order remainder function
for the central value of the renormalization scale $\mu_R=Q$
is provided
by the following function:
\bea
\label{eq_fit_rem3}
\mathcal{C_{\rm (fin.)}}(z) & \approx &
\, 15 \, \log^5(1-z)
\, + \, 130 \, \log^4(1-z)
\, + \, 408 \, \log^3(1-z) 
\nonumber\\
&+&  \, 544 \, \log^2(1-z)
\, + \, 308 \, \log(1-z) \, + \, 226 \, 
\\
&+&  \, 0.70545 \, \frac{ \log^2(z)}{z}
\, - \, 15.494 \, \frac{ \log (z)}{z}
\, + \, 39.568 \, \frac{1}{z}\,
\qquad
(0.01 \lsim z \lsim 0.98).
\nonumber
\eea
In order to improve the quality of the fit,
we have reduced the number of free parameters
by replacing the analytic values
of the coefficients of the terms enhanced for $z \to 0^+$ 
(the terms in the last line, containing the factor $1/z$).
The latter coefficients have been derived from a 
jet-calculus computation in Ref.\cite{Dixon:2019uzg}.

We have also fitted the first-order and
second-order remainder functions
with trial functions of analogous
form to (\ref{eq_fit_rem3}), finding in both cases
good agreement with the corresponding (exact) analytic functions
in all the kinematic range.

We have also performed the numerical calculation of the remainder functions in the case of resummation with the unitarity constraint of Eq.\,(\ref{Ltilde}):
\bea
\label{EEC-fintil}
\frac{1}{\sigma_{\rm tot}} \,\frac{ d \widetilde{\Sigma}_{\rm (fin.)}}{dz}
&=&
\mathcal{\widetilde{A}_{\rm (fin.)}}(z) \, \frac{\alpha_S}{\pi} 
 +  \mathcal{\widetilde{B}_{\rm (fin.)}}(z)  \left( \frac{\alpha_S}{\pi} \right)^2
 +  \mathcal{\widetilde{C}_{\rm (fin.)}}(z)  \left( \frac{\alpha_S}{\pi} \right)^3
 +   \mathcal{O}\left(\alpha_S^4\right)\,,
\eea
where (see Eqs.\,(\ref{EEC-exp},\ref{EEC-exptil},\ref{EEC-fin},\ref{EEC-fin2}))
\bea
\mathcal{\widetilde{A}_{\rm (fin.)}}(z)&=&\mathcal{{A}_{\rm (fin.)}}(z)+\mathcal{{A}_{\rm (res.)}}(z)-\mathcal{\widetilde{A}_{\rm (res.)}}(z)\,,\\
\mathcal{\widetilde{B}_{\rm (fin.)}}(z)&=&\mathcal{{B}_{\rm (fin.)}}(z)+\mathcal{{B}_{\rm (res.)}}(z)-\mathcal{\widetilde{B}_{\rm (res.)}}(z)\,,\\
\mathcal{\widetilde{C}_{\rm (fin.)}}(z)&=&\mathcal{{C}_{\rm (fin.)}}(z)+\mathcal{{C}_{\rm (res.)}}(z)-\mathcal{\widetilde{C}_{\rm (res.)}}(z)\,.
\eea
In this case the remainder functions
depend also on the resummation scale $\mu_Q$.  The fit for the third-order remainder function $\mathcal{\widetilde{C}_{\rm (fin.)}}(z)$ for $\mu_Q=\mu_R=Q$ is provided
by the following function:
\bea
\mathcal{\widetilde{C}_{\rm (fin.)}}(z) & \approx &
\, 15 \, \log^5(1-z)
\, + \, 130 \, \log^4(1-z)
\, + \, 408 \, \log^3(1-z) 
\nonumber\\
&+&  \, 548 \, \log^2(1-z)
\, + \, 304 \, \log(1-z) \, + \, 138 \,
\\
&+&  \, 0.70545 \, \frac{ \log^2(z)}{z}
\, - \, 15.494 \, \frac{ \log (z)}{z}
\, + \, 39.568 \, \frac{1}{z}\,
\qquad
(0.01 \lsim z \lsim 0.98).
\nonumber
\eea
We note that the (fitted) coefficients of the $\ln(1-z)$ terms are basically unchanged ($\mathcal{C_{\rm (fin.)}}(z)$ and $\mathcal{\widetilde{C}_{\rm (fin.)}}(z)$ have the same large-$z$ behavior), while only the constant term changes.

For reproducibility purposes we also provide below a fit of the
functions $\mathcal{\widetilde{A}_{\rm (fin.)}}(z)$ and
$\mathcal{\widetilde{B}_{\rm (fin.)}}(z)$ for $\mu_Q=\mu_R=Q$:
\bea
\mathcal{\widetilde{A}_{\rm (fin.)}}(z) & \approx &
\,\left(8.33 \,-\, \frac{12.43}z\right)\, \log(1-z)
\,-\,11.48  \,+\, 2.97\, z
\,+\, 1.05\, z^2 \,-\, 0.36\, z^3  \nn\\
&+& 0.80\, z^4 \,+\, \frac1{2z}\qquad
(0.01 \lsim z \lsim 0.99)\,;
\eea
\bea
\mathcal{\widetilde{B}_{\rm (fin.)}}(z) & \approx &
2.08\, \log^3(1-z) \,+\, 13.42\, \log^2(1-z)
\,+\, 38.28\, \log(1-z)
\,+\, 1.66  \nn\\
&+& 40.02\, z \,+\, 6.19\, z^2 \,+\, 1.35\, z^3
\,+\,  2.11\, z^4   
\,-\, 0.57\, \log(z) + \frac{4.59256}z \nn\\
&-& 0.720833\, \frac{\log(z)}z
\qquad
(0.01 \lsim z \lsim 0.99)\,.
\eea


\section{Numerical results}
\label{sec_scales}

In this section we apply the resummation formalism described in the previous
sections and we present some
illustrative numerical results for the EEC distribution. In particular
we present
perturbative predictions at NLL+LO, NNLL+NLO and N$^3$LL+NNLO
and we compare them with experimental data at the LEP accelerator.
We evaluate the QCD running coupling $\alpha_S(\mu_R^2)$
in the $\rm \overline{MS}$ renormalization
scheme at ($n+1$)-loop order at N$^n$LL accuracy with
$\alpha_S(m_Z^2) = 0.120$ and
we set the center--of--mass energy at the $Z^0$ peak ($Q=m_Z=91.1876\,$GeV).
We apply the unitarity constraint (see Eq.\,(\ref{Ltilde}) and related discussion)
such that our calculation exactly reproduces, 
after integration over the angular separation variable $\chi$, the corresponding fixed-order results for the total cross section of electron-positron annihilation into hadrons
up to N$^3$LO\,\cite{Gorishnii:1990vf,Surguladze:1990tg}.
In order to estimate the size of yet uncalculated higher-order terms
and the ensuing perturbative uncertainties we consider the dependence
of the results on the auxiliary scales $\mu_R$ and $\mu_Q$.
In particular we perform an independent variation of $\mu_R$ and $\mu_Q$ in the range
\beq
\label{muvar}
\frac{m_Z}{4} \, \leq \, \{\mu_R,\mu_Q\} \, \leq \, m_Z\,.
\eeq
with the constraint
\beq
\label{muvar2}
\frac{1}{2} \, \leq \, \mu_R/\mu_Q \, \leq \, 2\,.
\eeq

\begin{figure}[th]
\begin{center}
\includegraphics[width=.925\textwidth]{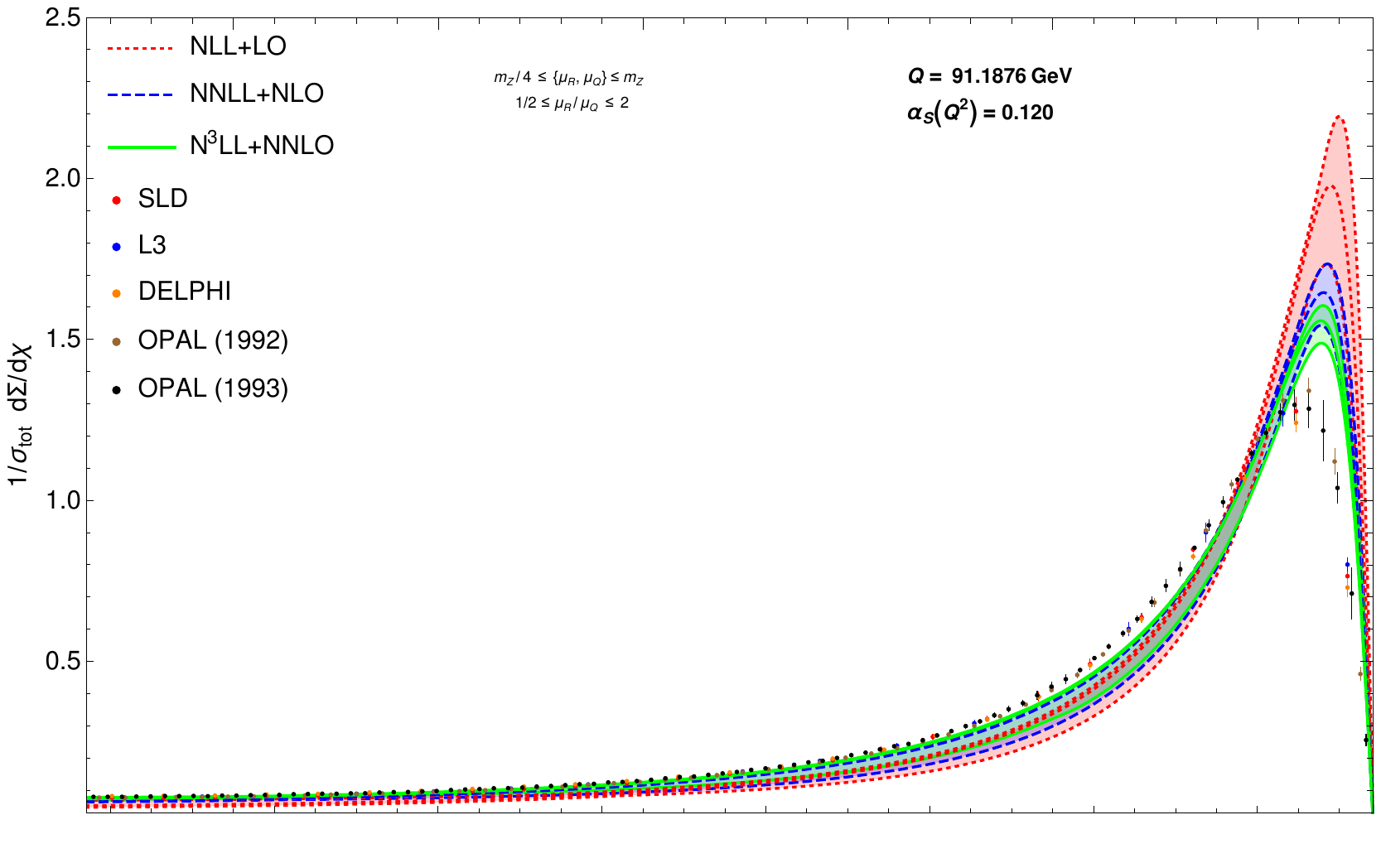}\vspace*{-.69cm}\\
\hspace*{-.035cm}\includegraphics[width=.924\textwidth]{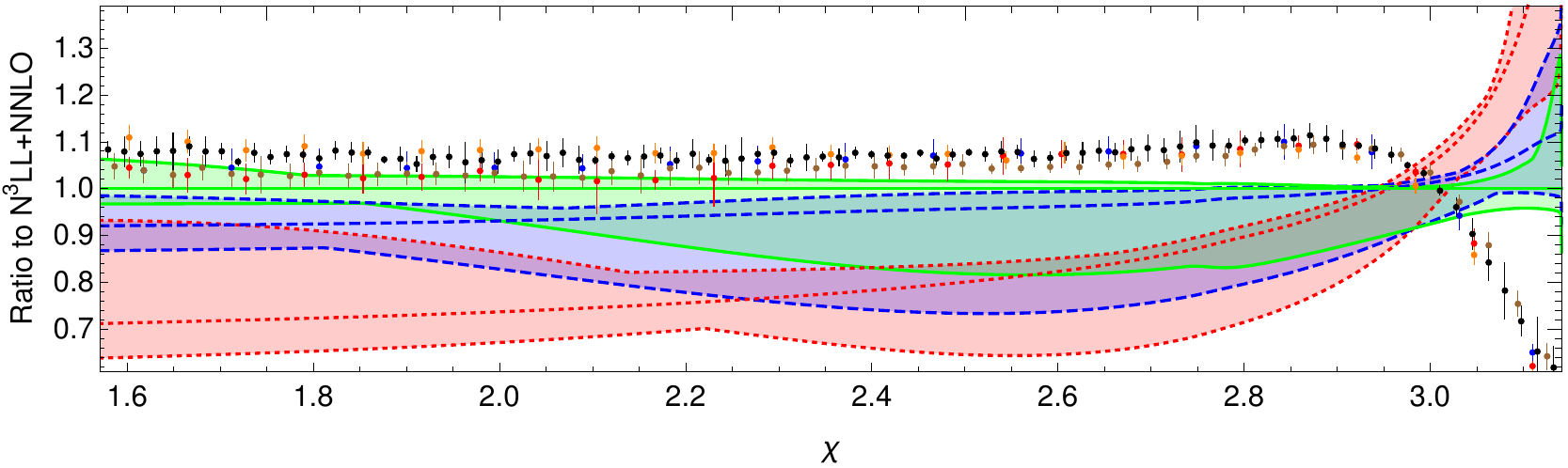}
\end{center}
\caption{
\label{figmain}
{\em
The EEC spectrum at $\sqrt{s}=91.1876$~GeV
at different perturbative orders in QCD.
}}
\end{figure}

The Landau singularity of the QCD coupling has been regularized
in a minimal way within the so-called
Minimal Prescription\,\cite{Catani:1996yz,Laenen:2000de,Kulesza:2002rh}.
We have also used the simpler procedure of integrating over the real $b$-axis with a sharp cut-off at a large value of
$b\lsim b_L$ or using the so-called ``$b^\star$ prescription''\,\cite{Collins:1981va,Collins:1984kg},  which smoothly freezes the integration over $b$ below a fixed upper limit $b\lsim b_L$. We found that the numerical
differences between the results obtained by these procedures are extremely small (i.e.\ much smaller than the perturbative uncertainties)
up to very large values of $\chi \to \pi$ ($\chi \sim 3.12$).

In Fig.\ref{figmain} we show the NLL+LO, NNLL+NLO and
N$^3$LL+NNLO predictions for the differential
distribution of the EEC function compared with experimental data from\,\cite{DELPHI:1992qrr,L3:1992btq,OPAL:1991uui,OPAL:1993pnw,SLD:1994idb}.
We observe that, by increasing the logarithmic accuracy,
the peak becomes smaller and broader  
and the low-$\chi$ tail becomes higher. This effect is not unexpected
since at higher perturbative accuracy subleading effects from parton
radiation are included.

In the lower panel of Fig.\,\ref{figmain}, we show the ratio of the predictions with respect to the N$^3$LL+NNLO
result at the central value of the scales $\mu_R=\mu_Q = Q/2$.
We observe that the scale dependence bands overlap in the entire angular region considered, thus
indicating that the bands represent a good estimate of  the perturbative uncertainty.
The alternative choice of $\mu_R=\mu_Q = Q$ as central value of the scales produces a
less conservative theoretical uncertainty\,\footnote{A similar behavior was observed and discussed in the closely related case of transverse-momentum
resummation in Sec. 3.1 of Ref.\,\cite{Catani:2015vma}.}.
The size of the N$^3$LL+NNLO (NNLL+NLO) scale dependence is about $9\%$ ($13 \%$) at $\chi \sim \pi/2$, then it increase up to 
$20\%$ ($27\%$) for $\chi \sim 2.5$, it reduces again down to $9\%$ ($17\%$) for $\chi \sim 3.1$ (i.e.\ around the peak) and it rapidly increases
in the $\chi \to \pi$ limit.
Overall going from NNLL+NLO to N$^3$LL+NNLO the scale uncertainty reduces by about a factor of 1.5-2.

From the results in Fig.\ref{figmain} we note that, by increasing the perturbative accuracy, the agreement with the 
experimental data improves, as expected.
However, the N$^3$LL+NNLO scale uncertainty band is definitively above
the experimental distribution in the peak region
and below it in the tail region.
This discrepancy calls
for  an explicit and consistent inclusion of non-perturbative effects of parton hadronization.

In Ref.\,\cite{Ebert:2020sfi} the resummation of the EEC function has been performed within the framework of SCET.
While a detailed numerical comparison is beyond our scope and capabilities, by looking at Fig.\,(3) of Ref.\,\cite{Ebert:2020sfi}
we observe that their N$^3$LL' prediction (which corresponds, in our notation, to N$^3$LL+NNLO accuracy) is qualitatively very similar to our result.
As in our case the N$^3$LL' curve of Ref.\,\cite{Ebert:2020sfi} slightly undershoots the experimental data in the region $2.6 \lesssim \chi \lesssim 3$ while
overshoots the data for $\chi \gtrsim  3$.

\begin{figure}[th]
\begin{center}
\includegraphics[width=.925\textwidth]{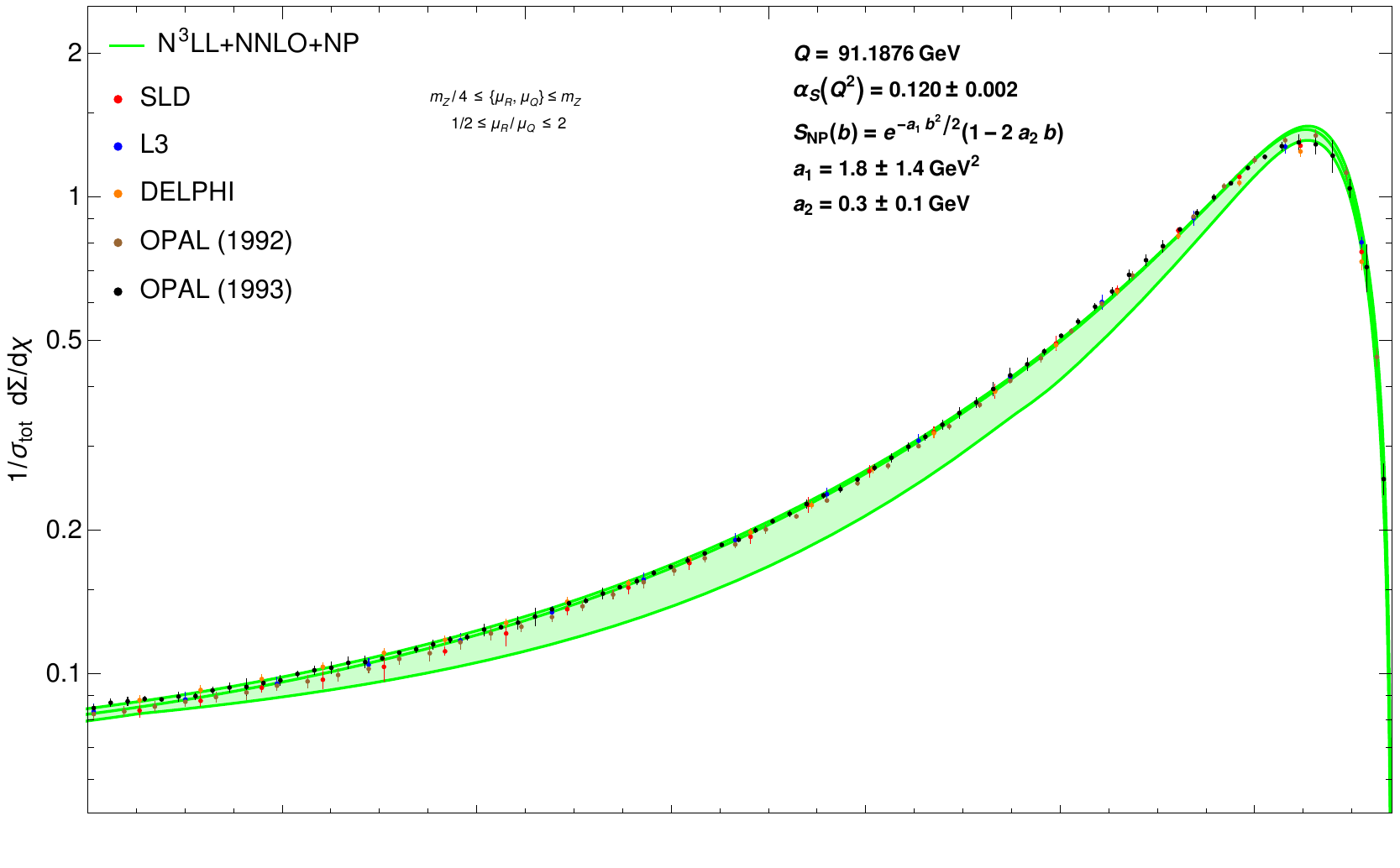}\vspace*{-.69cm}\\
\hspace*{-.035cm}\includegraphics[width=.924\textwidth]{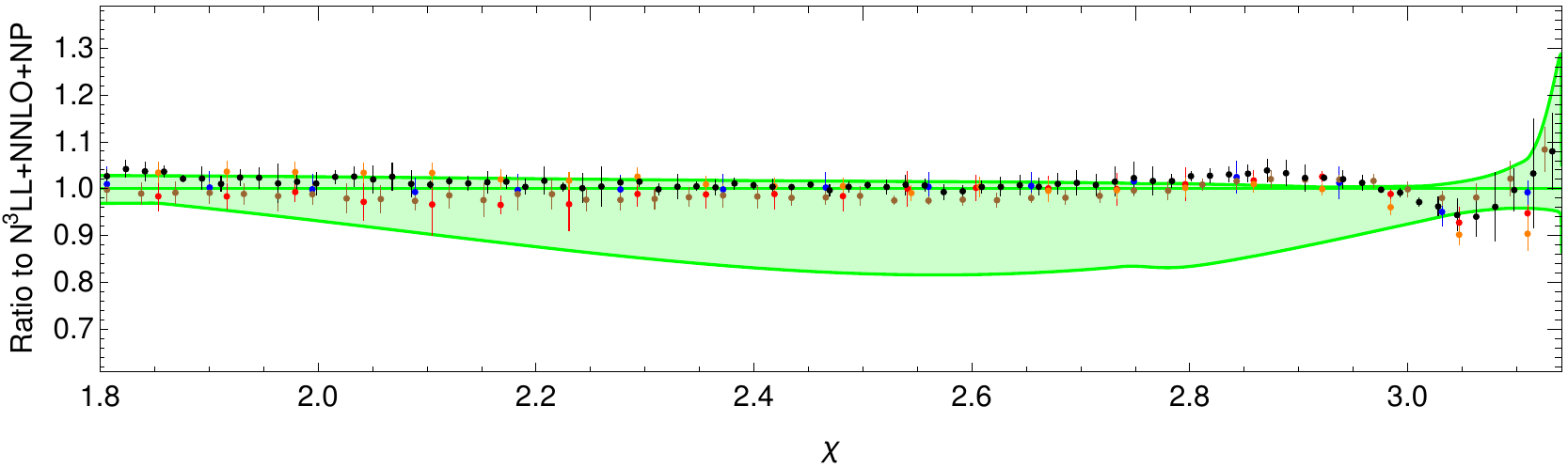}
\end{center}
\caption{
\label{figNP}
{\em
The EEC spectrum at $\sqrt{s}=91.1876$~GeV
at N$\,^3\!$LL+NNLO accuracy,
including QCD Non-Perturbative (NP) effects.
}}
\end{figure}


We finally extend
our perturbative results including  NP power-behaved QCD contributions
using the analytic ``dispersive approach'' of Ref.\,\cite{Dokshitzer:1995qm,Dokshitzer:1999sh}.
In such a model NP effects are introduced by multiplying the perturbative Sudakov form factor in Eq.\,\ref{sud2}
by a NP form factor, $S_{NP}$, of the form
\begin{equation}
\label{SNP}
S_{NP}(b) = e^{-\frac12 a_1 b^2}\left(1-2a_2 b\right)\,,
\end{equation}
where $a_1$ and $a_2$ are NP parameters that can be related to moments of an effective QCD coupling to be fitted to the data.

In this paper 
we have performed a preliminary three-parameter ($\alpha_S$, $a_1$ and $a_2$)
fit to experimental data from LEP and SLC accelerators\,\cite{DELPHI:1992qrr,L3:1992btq,OPAL:1991uui,OPAL:1993pnw,SLD:1994idb}
at the center-of-mass $\sqrt{s}=m_Z$.
In Figs.\,\ref{figNP} we compare the N$^3$LL+NNLO predictions at central value of the scales $\mu_R=\mu_Q=Q/2$
with the inclusion of NP effects fitted against experimental data (the errors in the fitted parameters include the experimental uncertainty only).
We have also shown the theory uncertainty from perturbative scales variation calculated as in Fig.\,\ref{figmain}.
We observe a nice agreement between theory and data over the entire region of the 
angular separation variable considered. A good agreement with experimental data in the central angular region was also obtained in Ref.\,\cite{Schindler:2023cww},
where NLO  QCD results were supplemented with NP effects through a renormalon analysis method.

\begin{figure}[th]
\begin{center}
\includegraphics[width=.925\textwidth]{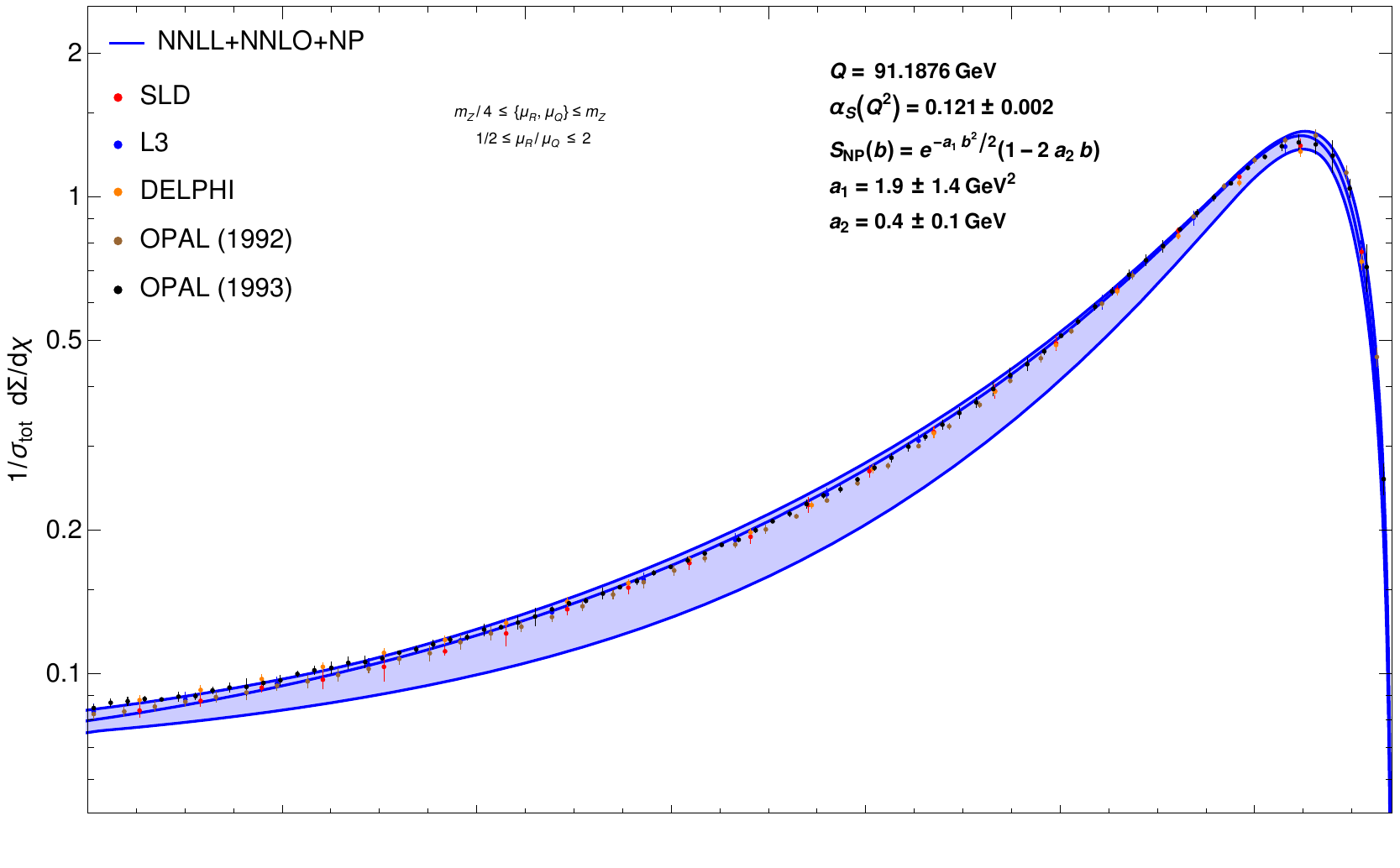}\vspace*{-.69cm}\\
\hspace*{-.02cm}\includegraphics[width=.923\textwidth]{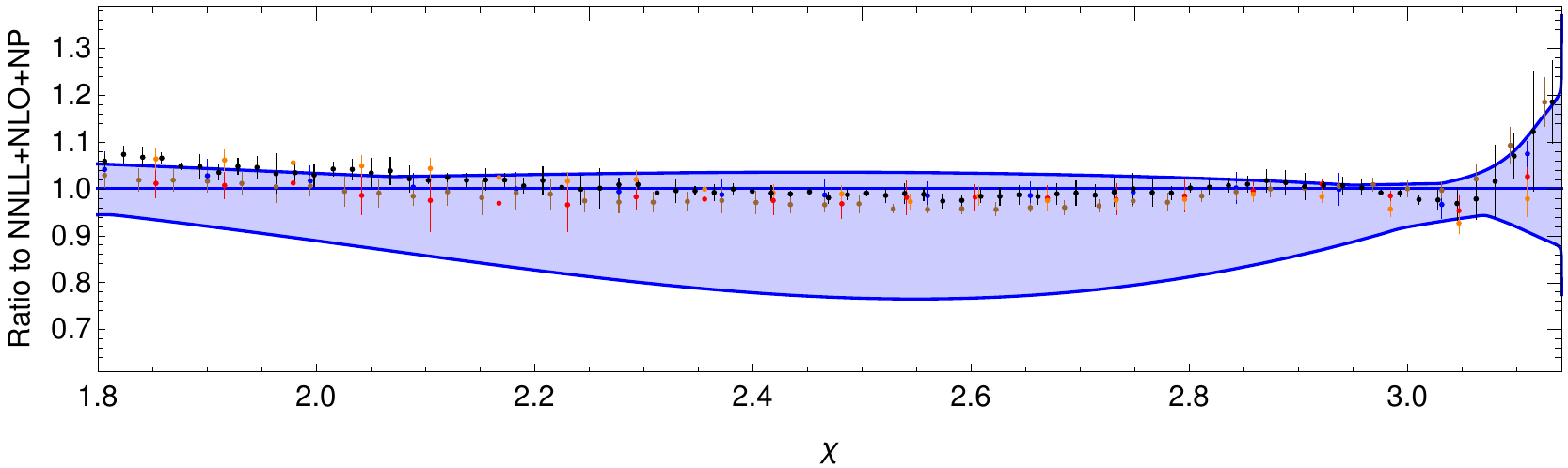}
\end{center}
\caption{
\label{figNP2} 
{\em
The EEC spectrum at $\sqrt{s}=91.1876$~GeV
at NNLL+NLO accuracy,
including QCD Non-Perturbative (NP) effects.
}}
\end{figure}

In order to test the validity of our results, we have performed the three-parameter fit also at NNLL+NLO.
The corresponding results are shown in Fig.\,\ref{figNP2}. 
A more detailed phenomenological analysis
with a comparison of experimental is left for a future work.

\section{Conclusions}
\label{sec_concl}

In this paper we have implemented the QCD
impact-parameter resummation formalism for the Energy-Energy-Correlation (EEC) distribution
in the back-to-back region, up to  next-to-next-to-next-to-leading logarithmic (N$^3$LL) accuracy.
Away from the back-to-back region, we have consistently combined resummed predictions with the known fixed-order
results up to next-to-next-to-leading order (NNLO). 

By expanding our QCD resummation formula up to second order in $\alpha_S$
and comparing the result with an exact QCD computation 
of the same order \cite{Ebert:2020sfi,Dixon:2018qgp}, we have been able to extract the
NNLO coefficient  of hard-virtual factor $H_2$.
The latter was the last missing piece for reaching the (full) next-to-next-to-leading logarithmic (NNLL) accuracy in QCD.
We have also been able to determine analytically the second-order
remainder function, which is relevant outside the
back-to-back region. 

By further expanding the QCD resummation formula
up to $\mathcal{O}(\alpha_S^3)$ and comparing with an 
analytic computation in the Soft-Collinear Effective Theory (SCET) 
in the back-to-back region \cite{Ebert:2020sfi},
we have also been able to extract the N$^3$LL  coefficient
$H_3$
of the hard-virtual factor (coefficient function),
as well as the N$^3$LL  coefficient $B_3$
of the single
logarithmic function of the Sudakov form factor.
This allowed us to perform
a complete resummation of the EEC function
in the back-to-back region at full N$^3$LL in QCD.
By subtracting the $\mathcal{O}(\alpha_S^3)$ 
part of the resummed distribution
from a third-order numerical computation of the EEC
in full QCD in the full angular range
($0<\chi<\pi$) \cite{Tulipant:2017ybb}, 
we have also been able to estimate the
NNLO (i.e.\ $\mathcal{O}(\alpha_S^3)$) remainder function.
By guessing the functional form of the third-order remainder
function, as well as using all the available
theoretical information, we have been able
to obtain a good fit for this function,
largely reducing the statistical uncertainties
of the single bins.
We have also checked the consistency between the
two available third-order calculations.
In particular, we have verified numerically
the SCET relation between the coefficient
$A_3$ in threshold and $q_\perp$ resummation 
\cite{Becher:2010tm}.

We have performed an independent variation of the  renormalization scale $\mu_R$ and the resummation scale $\mu_Q$   
as a way to estimate the perturbative theoretical error.
The scale variation band at N$^3$LL+NNLO is reduced by about a factor of 1.5-2 with respect to the previous order.

In the framework of a pure perturbative calculation, 
by going from NNLL+NLO to the N$^3$LL+NNLO accuracy, 
we find reasonably small corrections
indicating a good convergence of the perturbative expansion.

Finally, we have compared our perturbative predictions to the EEC distribution measured 
at the LEP and SLC accelerator at the $Z^0$ peak.
Increasing the perturbative accuracy improves the compatibility
of the theoretical predictions with the experimental data. 
However, there is still a substantial discrepancy between N$^3$LL+NNLO prediction and data.

Finally, after introducing, within a dispersive approach, non-perturbative power corrections,
we have been able to obtain an accurate description of experimental data at LEP and SLC accelerators
at the $Z^0$ peak.

A more detailed phenomenological analysis
is left for a  future work.

\vspace{1cm}
\centerline{\bf Acknowledgments}
~\\
\noindent
We would like to thank Gabor Somogyi for providing us the numerical results
of Ref.\,\cite{Tulipant:2017ybb}. 


\end{document}